\documentclass[12pt]{iopart}
\usepackage{color}
\usepackage{graphicx}
\usepackage{dcolumn}
\usepackage{bm}
\usepackage{amstext}
\usepackage{verbatim}
\usepackage{amssymb}
\usepackage[colorlinks,citecolor=blue,linkcolor=red,anchorcolor=blue,filecolor=blue,urlcolor=blue]{hyperref}
\usepackage{rotating}

\usepackage{iopams}

\begin{document}

\title[\underline{J. Phys. G: Nucl. Part. Phys. \hspace {6cm} Abdul Quddus {\it et al}}]
{Effective surface properties of light, heavy, and superheavy nuclei}


\author{Abdul Quddus$^{1}$, M. Bhuyan$^{2,3}$, and S. K. Patra$^{4,5}$}
\address{$^{1}$Department of Physics, Aligarh Muslim University, Aligarh-202002, India \\
$^{2}$Department of Physics, Faculty of Science, University of Malaya, Kuala Lumpur 50603, Malaysia \\
$^{3}$Institute of Research Development, Duy Tan University, Da Nang 550000 Vietnam \\
$^{4}$Institute of Physics, Bhubaneswar-751005, India \\
$^{5}$Homi Bhabha National Institute, Training School Complex, Anushakti Nagar, Mumbai 400085, India\\}
\date{\today}

\begin{abstract}
Starting from light to superheavy nuclei, we have calculated the effective surface properties such 
as the symmetry energy, neutron pressure, and symmetry energy curvature using the coherent density 
fluctuation model. The isotopic chains of O, Ca, Ni, Zr, Sn, Pb, and Z = 120 are considered in the 
present analysis, which cover nuclei over the whole nuclear chart. The matter density distributions 
of these nuclei along with the ground state bulk properties are calculated within the spherically 
symmetric effective field theory motivated relativistic mean field model by using the recently 
developed IOPB-I, FSUGarnet, and G3 parameter sets. The calculated results are compared with the 
predictions of the widely used NL3 parameter set and found in good agreement. We observe a few 
signatures of shell and/or sub-shell structure in the isotopic chains of nuclei. The present 
investigations are quite relevant for the synthesis of  exotic nuclei with high isospin asymmetry 
including superheavy and also to constrain an equation of state of nuclear matter. 
\end{abstract}


\maketitle

\section{Introduction} 
Nuclei lie away from $\beta-$ stability line with large neutron to proton asymmetry are of great 
importance. One of the quests among the nuclear physics community is how to synthesis the exotic 
and superheavy nuclei and to explore their applications. About 3000 nuclei, lie away from the 
$\beta$-stable line, have been synthesized in various laboratories and some more ($\sim 5000$) 
have to be synthesized. Since the matter at extreme density and temperature is impossible to create 
in a laboratory, a study of neutron-rich nuclei is treated as a tool to understand it. 
In the quest for the formation of superheavy nuclei, 
the last one with $Z=118$ has been synthesized at Dubna which was named Oganesson \cite{ogan06} 
and more superheavy nuclei are expected to synthesize. A lot of theoretical predictions are reported 
about the stability of superheavy nuclei against the spontaneous fission, $\alpha$- and $\beta$-decays, 
and neutron emission \cite{mehta,elena}. The mere existence of exotic nuclei including superheavy 
is entirely by the quantal shell effects, which play against the surface tension and the Coulomb 
repulsion. Furthermore, one of the very compelling issue in such exotic systems is the appearance 
of new magic numbers and the disappearance of others in moving from $\beta$-stable to drip-line 
region of the nuclear chart \cite{chou95,otsu01}. For example, beyond the proton number $Z=82$ and 
neutron number $N=126$, the next predicted magic numbers are Z = 114, 120, and 126 for the proton 
and N = 172 or 184 for the neutron \cite{elena}. The neutron-rich/deficient isotopes, and Z = 120 
element which is one of the predicted magic numbers represent a challenge for future experimental 
synthesis since they are located at the limit of accessibility with available cold fusion reactions 
facility. Therefore, an accurate estimation of their characteristics are essential from the theoretical 
side to guide future experiment. Various experiments around the globe like Jyav\"askyl\"a 
(Finland) \cite{jyav}, FRIB (US) \cite{frib}, GSI (Germany) \cite{gsi}, RIKEN (Japan) \cite{riken}, 
GANIL (France) \cite{ganil}, FLNR (Russia) \cite{flnr}, CSR (China) \cite{csr}, FAIR (Germany) 
\cite{fair}, and ORNL (US) \cite{ornl} provide a possibility of exploring an exotic nuclei, and superheavy nuclei
under extreme condition of isospin asymmetry. 
  
The nuclear symmetry energy is directly connected with the isospin asymmetry of the system; either 
infinite nuclear matter or finite nuclear system.
It is an important quantity having significant role in different areas of 
nuclear physics, for example, in structure of ground state nuclei \cite{Niksic08,Van10,Dalen10}, physics 
of giant collective excitation \cite{Rodin07}, dynamics of heavy-ion reactions \cite{Chen08,Colonna09}, 
and physics of neutron star \cite{Steiner05,fattoyev12,dutra12,dutra14}. It determines various neutron 
star properties such as mass-radius trajectory, its cooling rates, the thickness of the crust, and the 
moment of inertia \cite{lrp}. The astrophysical observations and availability of exotic beam in a 
laboratory have raised an interest in symmetry energy \cite{gai11}. In the last three decades, 
the density dependence of nuclear symmetry energy has played a great role to understand nuclei near the 
drip-line \cite{wd80}. For interpreting the neutron-rich 
nuclei and the neutron star matter, the characterization of the symmetry energy through experiments is 
a crucial step. But, the symmetry energy is not a directly measurable quantity. It is extracted from 
the observables related to it. Danielewicz has demonstrated that the ratio of the bulk symmetry energy 
to the surface symmetry energy is related to the neutron skin-thickness \cite{daniel03}. It is found 
that the radius of a neutron star is correlated to the density dependence of the symmetry energy at 
a saturation point \cite{lattimer07}. Furthermore, the $L$ coefficient (or say, pressure $P$) is correlated 
with the neutron skin-thickness of $^{208}$Pb \cite{brown,rj02,cent09,rocaprl11} and the radius of a 
neutron star. Even the precise measurement of neutron skin-thickness is difficult, yet \cite{prex1} 
it is one of the sensitive probes for nuclear symmetry energy. In some of our previous works 
as well as the works of others \cite{gai11,bhu18}, it has been shown that the symmetry energy of finite nuclei can be used 
as an observable to indicate/determine magic nuclei within Coherent density fluctuation model.

The neutron pressure of finite nuclei is related to the slope parameter ($L-$) of 
symmetry energy at saturation, which is an essential quantity in determining the equation of state (EoS) 
of nuclear matter \cite{gai11,bhu18,latt14,aqsubm}. Furthermore, for finite nuclei, the pressure depends 
on the strength of interaction among nucleons and their distributions. In our previous work 
\cite{aqsubm}, we have shown a correlation between neutron pressure and the neutron skin-thickness 
of neutron-rich thermally fissile nuclei. The symmetry energy and neutron pressure are collectively 
termed as effective surface properties, which are extensively defined in Refs. 
\cite{gai11,bhu18} and also illustrated in Sub-sec. \ref{denwef}. The importance of the surface 
properties and their sensitivity to density have motivated us to pursue their systematic study over 
all regions of the nuclear chart. Here, we have investigated the effective surface properties for the 
isotopic series of O, Ca, Ni, Zr, Sn, Pb, and Z = 120 nuclei, which cover all the region of the mass 
table from light to superheavy region. Recently, the symmetry energy of finite nuclei at a local 
density has been studied by using various formulae of the liquid drop model \cite{wd66,moller95,kp03}, 
the energy density functional of Skyrme force \cite{chen05,yosh06,chen10}, the random phase 
approximation based on the Hartree-Fock (HF) approach \cite{carbone10}, the relativistic 
nucleon-nucleon interaction \cite{lee98,bka10}, and the effective relativistic Lagrangian with 
density-dependent meson-nucleon vertex function \cite{dv03}. In Refs. 
\cite{gai11,bhu18,anto,gai12} the surface properties of the nuclei have been studied by 
folding the nuclear matter properties, within the Brueckner energy density functional 
\cite{brueck68,brueck69}, with the weight functions of the nuclei in the coherent density fluctuation 
model (CDFM) \cite{anto,antob}. The advantages of CDFM over other methods are that this method takes 
care (i) the fluctuation arises in the nuclear density distribution via weight function $\vert f (x) 
\vert^2$, and (ii) the momentum distributions through the mixed density matrix (i.e., the Wigner 
distribution function) \cite{bhu18,gai11,gai12}. In other words, the CDFM approach is 
adept to comprise the variation arise from the density and momentum distributions at the surface of 
finite nuclei. The present investigation covers systematic studies of the effective surface properties 
of several nuclei over the nuclear chart by finding the bulk properties along with densities of the 
nuclei within the effective field theory motivated relativistic mean field (E-RMF) approach. The 
calculated densities of the nuclei are served as the input to the CDFM to investigate the surface 
properties.

The paper is organized as follows: in Sec. \ref{theory} we present the formalism followed to carry 
out this work. In Sub-section \ref{RMF}, we outline effective field theory motivated relativistic 
mean field model, which has been used to calculate the ground state bulk properties and 
densities of the nuclei. Sub-section \ref{nuclear-matter} contains the general idea of calculating 
symmetry energy and relevant quantities like pressure and symmetry energy curvature. In Sub-section \ref{cdfm}
we present the formalism (CDFM) to calculate the effective surface properties. The calculated results 
are discussed in Sec. \ref{result}. Finally, the work is summarized in Sec. \ref{summary}.  

\section{Formalism}{\label{theory}}
\subsection {Effective field theory motivated relativistic mean field model (E-RMF)}
\label{RMF}
Relativistic mean field (RMF) theory is one of the microscopic approaches to solve the many body 
problem of nuclear system. In the RMF model, the nucleons are assumed to interact through the 
exchange of mesons. The model predicts ground as well as an intrinsic excited state properties of 
nuclei such as the binding energy, root mean square (RMS) radius, nuclear density distributions, 
deformation parameter, and single particle energies throughout the nuclear landscape. The 
details of the RMF models and their parameterizations can be found in Refs. 
\cite{furnstahl97,patra01,aqjpg,chai15,iopb1,G3,lala97}. For the sake of completeness, here we 
present the E-RMF formalism briefly. Effective mean field approximated Lagrangian density has, in 
principle, several numbers of terms with all possible types of self and cross-couplings of mesons. 
To handle E-RMF numerically, the ratios of fields and the nucleon mass are used in the truncation scheme 
as a constrain of naturalness. In this work, we have used the E-RMF Lagrangian having contributions 
of $\delta -$ meson and photon up to 2$^{th}$ order exponent and the rest up to 4$^{th}$ order of 
exponents, which has been shown to be reasonably good approximation to predict the finite nuclei 
and the nuclear matter observables up to considerable satisfaction \cite{G3}. The energy density, 
obtained within the E-RMF Lagrangian by applying mean field approximation, is given as: 
%
\begin{eqnarray}
{\cal E}({r}) & = &  \sum_i  \varphi_i^\dagger({r})
\Bigg\{ -i \mbox{\boldmath$\alpha$} \!\cdot\! \mbox{\boldmath$\nabla$}
+ \beta \left[M - \Phi (r) - \tau_3 D(r)\right] + W({r})
+ \frac{1}{2}\tau_3 R({r})
\nonumber \\[3mm]
& &
+ \frac{1+\tau_3}{2} A ({r})
- \frac{i \beta\mbox{\boldmath$\alpha$}}{2M}\!\cdot\!
  \left (f_\omega \mbox{\boldmath$\nabla$} W({r})
  + \frac{1}{2}f_\rho\tau_3 \mbox{\boldmath$\nabla$} R({r}) \right)
  \Bigg\} \varphi_i (r)
\nonumber \\[3mm]
& & \null
  + \left ( \frac{1}{2}
  + \frac{\kappa_3}{3!}\frac{\Phi({r})}{M}
  + \frac{\kappa_4}{4!}\frac{\Phi^2({r})}{M^2}\right )
   \frac{m_s^2}{g_s^2} \Phi^2({r})
- \frac{\zeta_0}{4!} \frac{1}{ g_\omega^2 } W^4 ({r})
\nonumber \\[3mm]
& & \null 
+ \frac{1}{2g_s^2}\left( 1 +
\alpha_1\frac{\Phi({r})}{M}\right) \left(
\mbox{\boldmath $\nabla$}\Phi({r})\right)^2
 - \frac{1}{2g_\omega^2}\left( 1 +\alpha_2\frac{\Phi({r})}{M}\right)
 \left( \mbox{\boldmath $\nabla$} W({r})  \right)^2
 \nonumber \\[3mm]
 & &  \null 
 - \frac{1}{2}\left(1 + \eta_1 \frac{\Phi({r})}{M} +
 \frac{\eta_2}{2} \frac{\Phi^2 ({r})}{M^2} \right)
  \frac{m_\omega^2}{g_\omega^2} W^2 ({r})
   - \frac{1}{2e^2} \left( \mbox{\boldmath $\nabla$} A({r})\right)^2
    \nonumber \\[3mm]
  & & \null
   - \frac{1}{2g_\rho^2} \left( \mbox{\boldmath $\nabla$} R({r})\right)^2
   - \frac{1}{2} \left( 1 + \eta_\rho \frac{\Phi({r})}{M} \right)
   \frac{m_\rho^2}{g_\rho^2} R^2({r})
   -\Lambda_{\omega}\left(R^{2}(r)\times W^{2}(r)\right)
    \nonumber \\[3mm]
  & & \null
   +\frac{1}{2 g_{\delta}^{2}}\left( \mbox{\boldmath $\nabla$} D({r})\right)^2
   +\frac{1}{2}\frac{ {  m_{\delta}}^2}{g_{\delta}^{2}}\left(D^{2}(r)\right)\;,
\label{eq1}
\end{eqnarray}
%
where $\Phi$, $W$, $R$, $D$ and $A$ are the fields which have been redefined as $\phi = g_\sigma 
\sigma$, $W = g_\omega \omega^0$, $R = g_\rho \rho^0$, and $A = e A^0$. $m_\sigma$, $m_\omega$, 
$m_\rho$ and $m_\delta$ are the masses and $g_\sigma$, $g_\omega$, $g_\rho$, $g_\delta$, 
$\frac{e^2}{4\pi}$ are the coupling constants for $\sigma$, $\omega$, $\rho$, $\delta$ mesons and 
photon, respectively. 
Using the Euler-Lagrangian equation of motion to Eq. \ref{eq1}, we get two first-order coupled differential 
equations for nucleons and four second-order differential equations for the four types of the meson fields \cite{iopb1}. 
The detailed equations can be found in Refs. \cite{iopb1, G3}. We transformed the Dirac equation into a Shr$\ddot{o}$dinger-like 
form as it is done in \cite{iopb1} by eliminating the smaller component of the Dirac spinor. Then the equation 
is solved by following the procedure of Vautherin and Brink \cite{brink72} with a fourth-order Runge-Kutta 
algorithm and the meson fields are solved by the Newton-Raphson method \cite{patra01,iopb1, G3}.

The total energy of a nucleus is given by following expression:
\begin{eqnarray}
E = \int {\cal E}(r) d^3r + E_{cm}+E_{pair};
\end{eqnarray}
where, first term has the contribution of mesonic and nucleonic energy densities given by Eq. \ref{eq1}. 
The second and third terms are the centre-of-mass correction energy and pairing energy, respectively. 
The expression for $E_{cm}$ is given as:
\begin{eqnarray}
E_{cm} &=& -\frac{3}{4} \times 41A^{-1/3}.
\end{eqnarray}

To describe open-shell nuclei (other than double magic nuclei), certainly pairing plays a crucial role. In general, 
the BCS approximation with a constant gap/force scheme is adopted on top of the RMF/E-RMF formalism to take care of 
the pairing correlation. However, this prescription does not hold good for drip-line nuclei as the seniority 
pairing recipe fails for such exotic nuclei. This is because the coupling to the continuum in the normal BCS 
approximation is not taken correctly. This deficiency can be removed to some extent by including few quasi-bound 
states owing to their centrifugal barrier \cite{chabanat98}. These quasi-bound states mock up the influence of 
the continuum.  Here, we follow the procedure of Refs. \cite{patra01}. For drip-line nuclei, there are no bound 
single-particle 
levels above the Fermi surface. In our calculations, we take the bound-state contributions and the levels coming 
from the quasi-bound states at positive energies \cite{chabanat98} and the expressions for $E_{pair}$ is written as:
\begin{eqnarray}
E_{pair} &=& -\frac{\Delta_i^2}{G_i} , \\ \nonumber
\end{eqnarray}
where $\Delta_i$ and $G_i (= C_i/A)$ are, respectively, the pairing gap and strength with $i=n, p$. 
The $C_i$ are chosen in a way to reproduce the binding energy of a nucleus with mass number $A$. 
For G2 set, $C_n = 21$ and $C_p = 22.5$ MeV, and for FSUGarnet, G3 and IOPB-I sets, $C_n = 19$ and $C_p = 21$ MeV.

\subsection{The key equation of state parameters in nuclear matter}
\label{nuclear-matter}
The total energy of a nucleus given in the liquid droplet model, which is an extension of the 
Bethe Weizsãcker liquid drop model incorporating the volume and surface asymmetry is expressed as \cite{Steiner05,wd66}: 
\begin{eqnarray}
E(A, Z) &=& -B.A + E_S A^{2/3} + S_V A \frac{(1 - 2Z/A)^2}{(1 + S_S A^{-1/3}/S_V)} + E_C \frac{Z^2}{A^{1/3}} 
\nonumber \\[3mm]
&& \null 
+ E_{dif} \frac{Z^2}{A} + E_{ex} \frac{Z^{4/3}}{A^{1/3}} + a \Delta A^{-1/2}.
\end{eqnarray}
Where $B\sim 16$ MeV is the binding energy per nucleon of bulk symmetric matter at saturation. $E_S$, $E_C$, $E_{dif}$, and $E_{ex}$ 
are the coefficients for the surface energy of symmetric matter, the Coulomb energy of a uniformly charged sphere, 
the diffuseness correction, and the exchange correction to the Coulomb energy, respectively. The last term gives the pairing 
corrections, which is essential for open-shell nuclei. $S_V$ is the volume symmetry energy parameter and $S_S$ is the modified 
surface symmetry energy parameter in the liquid model (see Ref. \cite{Steiner05}). The third term on the right-hand side of the above 
equation is the contribution of the symmetry energy to the total energy of a nucleus which can not be neglected and has a significant 
impact where N/Z ratio widely differs from the value at the valley of stability (drip-line nuclei). 

Moreover, the energy per nucleon of nuclear matter ${\cal E}/A$=$e(\rho, \alpha)$ (where $\rho$ is the baryon 
density) can be expanded by Taylor series expansion method in terms of isospin asymmetry parameter 
$\alpha\left(=\frac{\rho_n-\rho_p}{\rho_n+\rho_p}\right)$:
\begin{eqnarray}
e(\rho, \alpha)=\frac{{\cal E}}{\rho_{B}} - M =
{e}(\rho) + S(\rho) \alpha^2 + 
{\cal O}(\alpha^4) ,
\end{eqnarray}
where ${e}(\rho)$, $S(\rho)$ and $M$ are the energy density of symmetric nuclear matter (SNM) ($\alpha$ 
= 0), the symmetry energy, and the mass of a nucleon, respectively. The odd powers of $\alpha$ are 
forbidden by the isospin symmetry and the terms proportional to $\alpha^4$ and higher orders have a negligible 
contribution. The symmetry energy $S(\rho)$ is defined by:
\begin{eqnarray}
S(\rho)=\frac{1}{2}\left[\frac{\partial^2 {e}(\rho, \alpha)}
{\partial \alpha^2}\right]_{\alpha=0}.
\label{s0v}
\end{eqnarray}
Near the saturation density $\rho_0$, the symmetry energy can be expanded through the Taylor series 
expansion method as:
\begin{eqnarray}
S(\rho)=J + L{\cal Y} 
+ \frac{1}{2}K_{sym}{\cal Y}^2 +\frac{1}{6}Q_{sym}{\cal Y}^3 + {\cal O}[{\cal Y}^4],
	\label{eq30}
\end{eqnarray} 
where $J=S(\rho_0)$ is the symmetry energy at saturation and ${\cal Y} = \frac{\rho-\rho_0}{3\rho_0}$. 
The slope parameter ($L -$ coefficient), the symmetry energy curvature ($K_{sym} $), and the skewness 
parameter ($Q_{sym}$) are defined as:
\begin{eqnarray}
	L=3\rho\frac{\partial S(\rho)}{\partial {\rho}}\bigg{|}_{ \rho=\rho_0},\;
\end{eqnarray}
\begin{eqnarray}
	K_{sym}=9\rho^2\frac{\partial^2 S(\rho)}{\partial {\rho}^2}\bigg{|}_{ {\rho=\rho_0}},\;
\end{eqnarray}
\begin{eqnarray}
        Q_{sym}=27\rho^3\frac{\partial^3 S(\rho)}{\partial {\rho}^3}\bigg{|}_{ {\rho=\rho_0}},\;
\label{qsym}
\end{eqnarray}
respectively. The neutron pressure of asymmetric nuclear matter can also be evaluated from the slope 
parameter ($L -$) by using the relation: 
\begin{eqnarray}
L^{NM}=\frac{3 P_0^{NM}}{\rho_0}.
\end{eqnarray}
It is to note that these nuclear matter properties at saturation (i.e., $\rho_0$, $S(\rho_0)$, 
$L(\rho_0)$, $K_{sym}(\rho_0)$, and $Q_{sym}(\rho_0)$) are model dependent and vary with certain 
uncertainties. More details of these quantities and their values along with the 
allowed ranges for the non-relativistic and relativistic mean field models with various force 
parameters can be found in Refs. \cite{dutra12,dutra14}. 

\subsection{The coherent density fluctuation model (CDFM)}
\label{cdfm}
The CDFM was suggested and developed in Refs. \cite{anto,antob}. In the CDFM, the one-body density 
matrix $\rho$ ({\bf r, r$'$}) of a nucleus can be written as a coherent superposition of infinite 
number of one-body density matrices $\rho_x$ ({\bf r}, {\bf r$'$}) for spherical pieces of the nuclear 
matter called as {\it Fluctons} \cite{gai11,bhu18,anto},
\begin{equation}
\rho_x ({\bf r}) = \rho_0 (x) \Theta (x - \vert {\bf r} \vert), 
\label{denx} 
\end{equation}
with $\rho_o (x) = \frac{3A}{4 \pi x^3}$. The generator coordinate $x$ is the spherical radius of 
the nucleus contained in a uniformly distributed spherical Fermi gas. In finite nuclear system, the 
one body density matrix can be given as \cite{gai11,bhu18,anto,gai12},
\begin{equation}
\rho ({\bf r}, {\bf r'}) = \int_0^{\infty} dx \vert f(x) \vert^2 \rho_x 
({\bf r}, {\bf r'}), 
\label{denr} 
\end{equation}
where, $\vert f(x) \vert^2 $ is the weight function (Eq. (\ref{wfn})). The term $\rho_x ({\bf r}, 
{\bf r'})$ is the coherent superposition of the one body density matrix and defined as,
\begin{eqnarray}
\rho_x ({\bf r}, {\bf r'}) &=& 3 \rho_0 (x) \frac{J_1 \left( k_f (x) \vert 
{\bf r} - {\bf r'} \vert \right)}{\left( k_f (x) \vert {\bf r} - {\bf r'} 
\vert \right)} 
\times \Theta \left(x-\frac{ \vert {\bf r} + {\bf r'} \vert }{2} \right). 
\label{denrr}
\end{eqnarray}
Here, J$_1$ is the first order spherical Bessel function. The Fermi momentum of nucleons in the 
Fluctons with radius $x$ is expressed as $k_f (x)=(3\pi^2/2\rho_0(x))^{1/3} =\gamma/x$, where 
$\gamma=(9\pi A/8)^{1/3}\approx 1.52A^{1/3}$. The Wigner distribution function of the one body 
density matrices in Eq. (\ref{denrr}) is,
\begin{eqnarray}
W ({\bf r}, {\bf k}) =  \int_0^{\infty} dx \vert f(x) \vert^2 W_x ({\bf r}, {\bf k}). 
\label{wing}
\end{eqnarray}
Here, $W_x ({\bf r}, {\bf k})=\frac{4}{8\pi^3}\Theta (x-\vert {\bf r} \vert)\Theta (k_F(x)-\vert 
{\bf k} \vert)$. 
Similarly, the density $\rho$ (r) within CDFM can express in terms of the same weight function as,
\begin{eqnarray}
\rho (r) &=& \int d{\bf k} W ({\bf r}, {\bf k}) 
 = \int_0^{\infty} dx \vert f(x) \vert^2 \frac{3A}{4\pi x^3} \Theta(x-\vert 
{\bf r} \vert)
\label{rhor}
\end{eqnarray}
and it is normalized to the nucleon numbers of the nucleus, $\int \rho ({\bf r})d{\bf r} = A$. By 
taking the $\delta$-function approximation to the Hill-Wheeler integral equation, we can obtain the 
differential equation for the weight function in the generator coordinate \cite{anto,bhu18}. The 
weight function for a given density distribution $\rho$ (r) can be expressed as,
\begin{equation}
|f(x)|^2 = - \left (\frac{1}{\rho_0 (x)} \frac{d\rho (r)}{dr}\right )_{r=x}, 
\label{wfn}
\end{equation}
with $\int_0^{\infty} dx \vert f(x) \vert^2 =1$. For a detailed analytical derivation, one can follows 
Refs. \cite{bhu18,anto94,fuch95}. The symmetry energy, neutron pressure, and symmetry energy 
curvature for a finite nucleus are defined below by weighting the corresponding quantities for 
the infinite nuclear matter within the CDFM. The CDFM allows us to make a transition from the properties 
of nuclear matter to those of finite nuclei. Following the CDFM approach, the expression for the 
effective symmetry energy $S$, pressure $P$, and curvature $\Delta K$ for a nucleus can be written 
as \cite{gai11,bhu18,gai12,anto94,fuch95,anto17},
\begin{eqnarray}
S = \int_0^{\infty} dx \vert f(x) \vert^2 S_0^{NM} (\rho (x)) ,
\label{s0}
\end{eqnarray}
\begin{eqnarray}
P =  \int_0^{\infty} dx \vert f(x) \vert^2 P_0^{NM} (\rho (x)), 
\label{p0}
\end{eqnarray}
\begin{eqnarray}
\Delta K =  \int_0^{\infty} dx \vert f(x) \vert^2 \Delta K_0^{NM} (\rho (x)).  
\label{k0}
\end{eqnarray}
Here, the quantities on the left-hand-side of Eqs. (\ref{s0}-\ref{k0}) are the surface weighted average 
of the corresponding nuclear matter quantities with local density approximation, which have been 
determined within the method of Brueckner {\it et. al.}, \cite{brueck68,brueck69}.

In the present work considering the pieces of nuclear matter with density $\rho_0$(x), we have used 
the matrix element V(x) of the nuclear Hamiltonian the corresponding energy of nuclear matter from 
the method of Brueckner {\it et. al.}, \cite{brueck68,brueck69}. In Brueckner energy density functional 
method, the V(x) is given by:
\begin{eqnarray}
V(x)=A V_0(x) + V_C + V_{CO},
\label{vx}
\end{eqnarray}
where
\begin{eqnarray}
V_0(x)=37.53 [(1+\delta)^{5/3}+(1-\delta)^{5/3}]\rho_0^{2/3}(x) 
+b_1 \rho_0(x) +b_2 \rho_0^{4/3}(x)  \nonumber \\[3mm]
+b_3\rho_0^{5/3}(x) + \delta^2[b_4 \rho_0 (x)+b_5 \rho_0 ^{4/3}(x) + b_6 \rho_0^{5/3}], 
\label{bruc}
\end{eqnarray}
with
$b_1=-741.28$, $b_2=1179.89$, $b_3=-467.54$, $b_4=148.26$, $b_5=372.84$, and $b_6=-769.57$. The 
$V_0$(x) in Eq. \ref{vx} is the energy per particle of nuclear matter (in MeV) which accounts for 
the neutron-proton asymmetry. $V_C$ is the coulomb energy of charge particle (proton) in a flucton,
\begin{eqnarray}
V_C=\frac{3}{5} \frac{Z^2 e^2}{x},
\end{eqnarray}
and $V_{CO}$ is the coulomb exchange energy given by
\begin{eqnarray}
V_{CO}=0.7386 Z e^2 (3 Z /4 \pi x^3)^{1/3}.
\end{eqnarray}

On substituting $V_0$(x) in Eq. \ref{s0v} and taking its second order derivative, the symmetry energy 
$S_0^{NM}$(x) of nuclear matter with density $\rho_0$(x) is obtained:
\begin{eqnarray}
S_0^{NM}(x) = 41.7 \rho_0^{2/3}(x) + b_4 \rho_0(x) + b_5 \rho_0^{4/3}(x) + b_6 \rho_0^{5/3} (x). 
\end{eqnarray}
The corresponding parameterized expressions for the pressure $P_0^{NM}$(x) and the symmetry energy 
curvature $\Delta K_0^{NM}$(x) for such a system within Brueckner energy density functional method 
have the forms
\begin{eqnarray}
P_0^{NM}(x) = 27.8 \rho_0^{5/3}(x) + b_4 \rho_0^2(x) + \frac{4}{3} b_5 \rho_0^{7/3}(x)+ \frac{5}{3} 
b_6\rho_0^{8/3}(x), 
\end{eqnarray}
and 
\begin{eqnarray}
\Delta K_0^{NM}(x) = -83.4 \rho_0^{2/3}(x) + 4 b_5 \rho_0^{4/3}(x)+ 10 b_6 \rho_0^{5/3}(x), 
\end{eqnarray}
respectively. These quantities are folded in the Eqs. (\ref{s0}-\ref{k0}) with the weight function to 
find the corresponding quantities of finite nuclei within the CDFM.

\begin{table}
\caption{The FSUGarnet \cite{chai15}, IOPB-I \cite{iopb1}, G3 \cite{G3}, and NL3 \cite{lala97} parameter sets 
are listed. The nucleon mass $M$ is 939.0 MeV in all the sets.  All the coupling constants are dimensionless, 
except $k_3$ which is in fm$^{-1}$. The lower panel of the table shows the nuclear matter properties at 
saturation density $\rho_{0}$ (fm$^{-3}$).}
\renewcommand{\tabcolsep}{0.25cm}
\renewcommand{\arraystretch}{1.25}
\begin{tabular}{cccccccccc}
\hline
\hline
\multicolumn{1}{c}{}
&\multicolumn{1}{c}{NL3}
&\multicolumn{1}{c}{FSUGarnet}
&\multicolumn{1}{c}{G3}
&\multicolumn{1}{c}{IOPB-I}\\
\hline
$m_{s}/M$  &  0.541  &  0.529&  0.559&0.533  \\
$m_{\omega}/M$  &  0.833  & 0.833 &  0.832&0.833  \\
$m_{\rho}/M$  &  0.812 & 0.812 &  0.820&0.812  \\
$m_{\delta}/M$   & 0.0  &  0.0&   1.043&0.0  \\
$g_{s}/4 \pi$  &  0.813  &  0.837 &  0.782 &0.827 \\
$g_{\omega}/4 \pi$  &  1.024  & 1.091 &  0.923&1.062 \\
$g_{\rho}/4 \pi$  &  0.712  & 1.105&  0.962 &0.885  \\
$g_{\delta}/4 \pi$  &  0.0  &  0.0&  0.160& 0.0 \\
$k_{3} $   &  1.465  & 1.368&    2.606 &1.496 \\
$k_{4}$  &  -5.688  &  -1.397& 1.694 &-2.932  \\
$\zeta_{0}$  &  0.0  &4.410&  1.010  &3.103  \\
$\eta_{1}$  &  0.0  & 0.0&  0.424 &0.0  \\
$\eta_{2}$  &  0.0  & 0.0&  0.114 &0.0  \\
$\eta_{\rho}$  &  0.0  & 0.0&  0.645& 0.0  \\
$\Lambda_{\omega}$  &  0.0  &0.043 &  0.038&0.024   \\
$\alpha_{1}$  &  0.0  & 0.0&   2.000&0.0  \\
$\alpha_{2}$  &  0.0  & 0.0&  -1.468&0.0  \\
$f_\omega/4$  &  0.0  & 0.0&  0.220&0.0 \\
$f_\rho/4$  &  0.0  & 0.0&    1.239&0.0 \\
$f_\rho/4$  &  0.0  & 0.0&    1.239&0.0 \\
$\beta_\sigma$  &  0.0  & 0.0& -0.087& 0.0  \\
$\beta_\omega$  &  0.0  & 0.0& -0.484& 0.0  \\
\hline
\hline
$\rho_{0}$ (fm$^{-3})$ &  0.148  &  0.153&  0.148&0.149  \\
$\mathcal{E}_{0}$(MeV)  &  -16.29  & -16.23 &  -16.02&-16.10  \\
$M^{*}/M$  &  0.595 & 0.578 &  0.699&0.593  \\
$J$(MeV)   & 37.43  &  30.95&   31.84&33.30  \\
$K_{\infty}$(MeV)&271.38&229.5&243.96&222.65\\
$L$(MeV)  &  118.65  &  51.04 &  49.31&63.58 \\
$K_{sym}$(MeV)  &  101.34  & 59.36 & -106.07&-37.09 \\
$Q_{sym}$(MeV)  &  177.90  & 130.93&  915.47 &862.70  \\
\hline
\hline
\end{tabular}
\label{table1}
\end{table}

\begin{table}
\caption{The calculated binding energy per particle (B/A), and charge radius ($R_{ch}$) are compared with the 
available experimental data \cite{audi12,angeli13}. The predicted neutron skin-thickness $\Delta r= R_n - 
R_p$ is also depicted with all the four models and compared with the available experimental data \cite{jast}.}
\begin{tabular}{cccccccccc}
\hline
\hline
\multicolumn{1}{c}{Nucleus}&
\multicolumn{1}{c}{Obs.}&
\multicolumn{1}{c}{Expt.}&
\multicolumn{1}{c}{NL3}&
\multicolumn{1}{c}{FSUGarnet}&
\multicolumn{1}{c}{G3}&
\multicolumn{1}{c}{IOPB-I}\\
\hline
\hline
$^{16}$O  & B/A             & 7.976 & 7.917 & 7.876 & 8.037 & 7.977  \\
          & R$_{ch}$        & 2.699 & 2.714 & 2.690 & 2.707 & 2.705  \\
          & R$_{n}$-R$_{p}$ & -     &-0.026 &-0.029 &-0.028 &-0.027  \\
\\
$^{28}$O  & B/A             & 5.988 & 6.379 & 5.933 & 6.215 & 6.220 \\
          & R$_{ch}$        & -     & 2.800 & 2.804 & 2.791 & 2.805 \\
          & R$_{n}$-R$_{p}$ & -     & 0.809 & 0.796 & 0.741 & 0.809 \\
\\
$^{40}$Ca & B/A             & 8.551 & 8.540 & 8.528 & 8.561 & 8.577  \\
          & R$_{ch}$        & 3.478 & 3.466 & 3.438 & 3.459 & 3.458  \\
          & R$_{n}$-R$_{p}$ &-0.08     &-0.046 &-0.051 &-0.049 &-0.049  \\
\\
$^{48}$Ca & B/A             & 8.666 & 8.636 & 8.609 & 8.671 & 8.638 \\
          & R$_{ch}$        & 3.477 & 3.443 & 3.426 & 3.466 & 3.446 \\
          & R$_{n}$-R$_{p}$ & 0.16  & 0.229 & 0.169 & 0.174 & 0.202  \\
\\
$^{68}$Ni & B/A             & 8.682 & 8.698 & 8.692 & 8.690 & 8.707  \\
          & R$_{ch}$        & -     & 3.870 & 3.861 & 3.892 & 3.873  \\
          & R$_{n}$-R$_{p}$ & -     & 0.262 & 0.184 & 0.190 & 0.223  \\
\\
$^{90}$Zr & B/A             & 8.709 & 8.695 & 8.693 & 8.699 & 8.691  \\
          & R$_{ch}$        & 4.269 & 4.253 & 4.231 & 4.276 & 4.253 \\
          & R$_{n}$-R$_{p}$ & 0.09  & 0.115 & 0.065 & 0.068 & 0.091 \\
\\
$^{100}$Sn& B/A             & 8.253 & 8.301 & 8.298 & 8.266 & 8.284 \\
          & R$_{ch}$        & -     & 4.469 & 4.426 & 4.497 & 4.464  \\
          & R$_{n}$-R$_{p}$ & -     &-0.073 &-0.078 &-0.079 &-0.077 \\
\\
$^{132}$Sn& B/A             & 8.355 & 8.371 & 8.372 & 8.359 & 8.352 \\
          & R$_{ch}$        & 4.709 & 4.697 & 4.687 & 4.732 & 4.706 \\
          & R$_{n}$-R$_{p}$ & -     & 0.349 & 0.224 & 0.243 & 0.287 \\
\\
$^{208}$Pb& B/A             & 7.867 & 7.885 & 7.902 & 7.863 & 7.870  \\
          & R$_{ch}$        & 5.501 & 5.509 & 5.496 & 5.541 & 5.521 \\
          & R$_{n}$-R$_{p}$ & 0.17  & 0.283 & 0.162 & 0.180 & 0.221\\
\hline
\hline
\end{tabular}
\label{table3}
\end{table}

\section{Results and discussion}
\label{result}

\begin{figure*}[!b]
        \includegraphics[width=0.33\columnwidth,height=5.0cm]{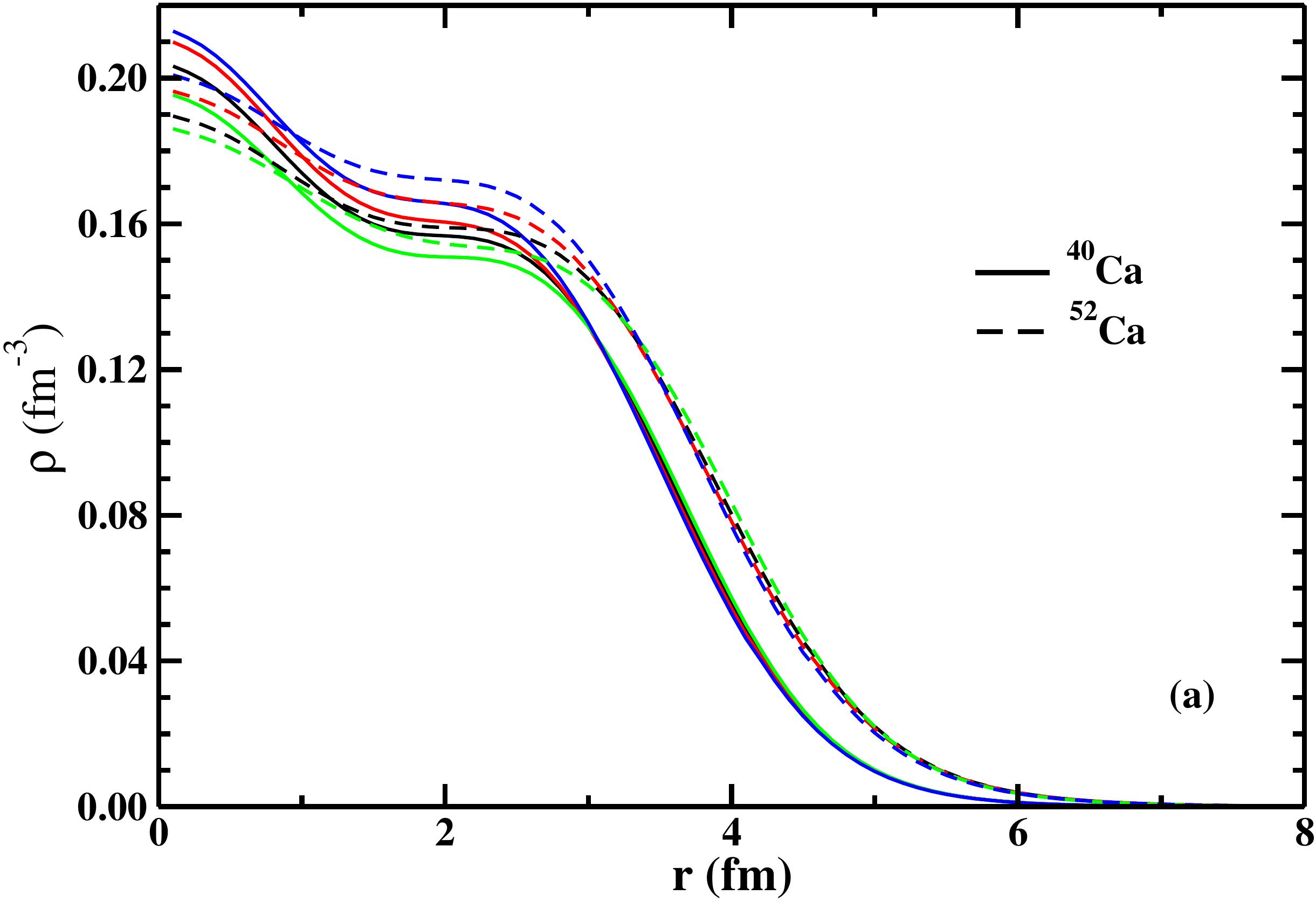}
        \includegraphics[width=0.33\columnwidth,height=5.0cm]{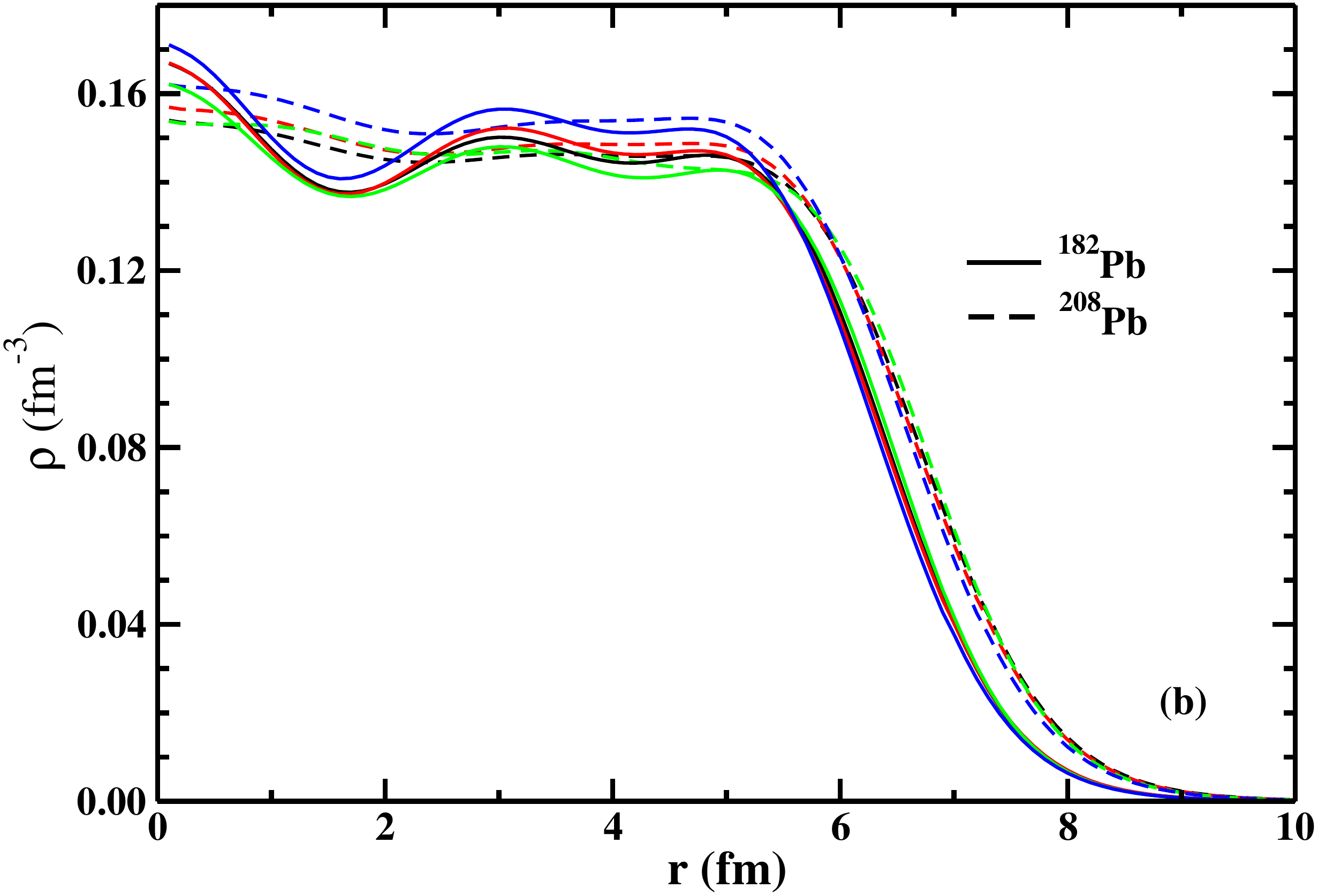}
        \includegraphics[width=0.33\columnwidth,height=5.0cm]{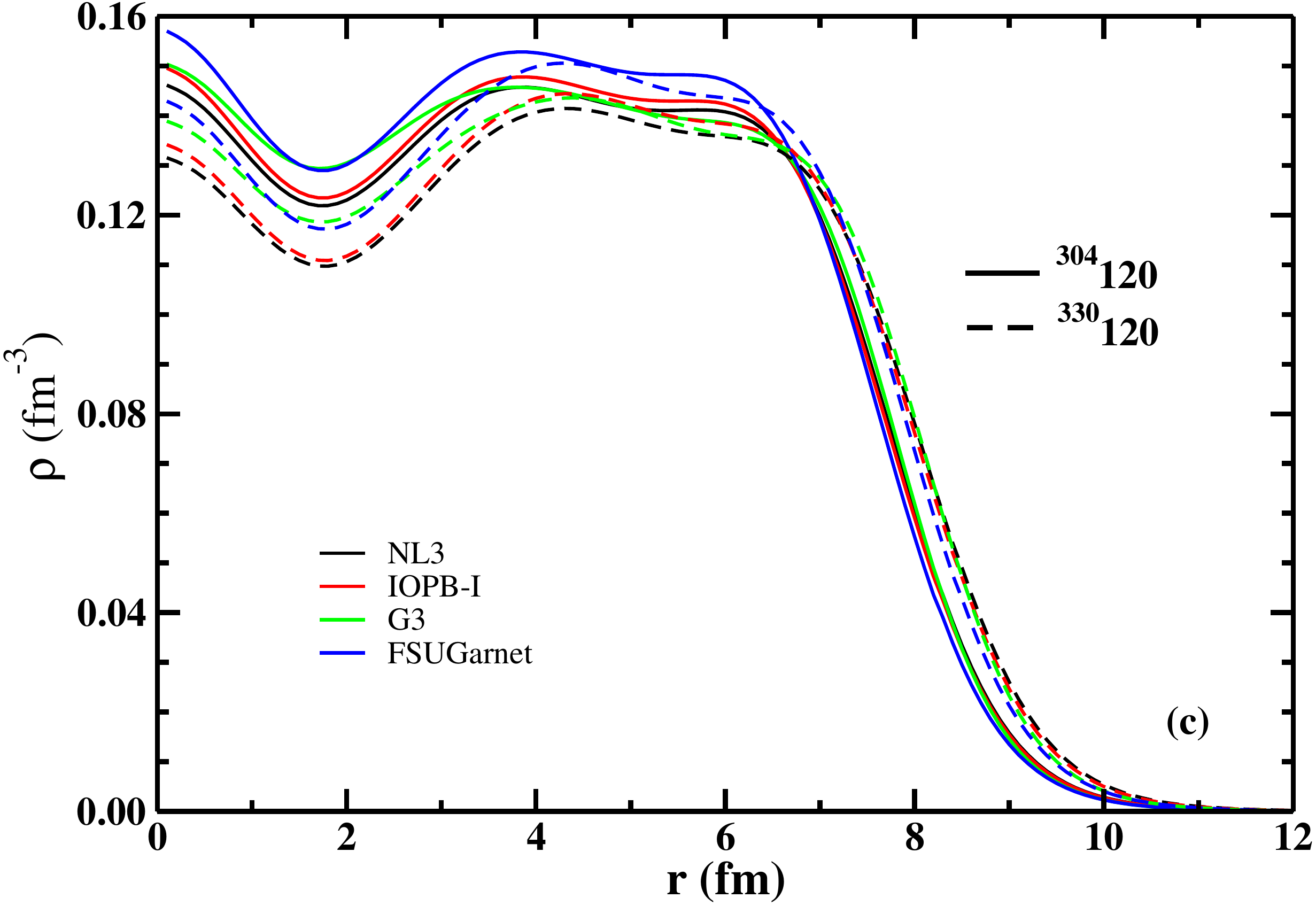}
\caption{(color online) (a) The total density profiles for (a) $^{40,52}$Ca, (b) $^{182,208}$Pb, and (c) 
$^{304,330}120$ as the representative cases corresponding to  FSUGarnet \cite{chai15}, IOPB-I \cite{iopb1}, 
G3 \cite{G3}, and NL3 \cite{lala97} parameter sets.}
\label{dens}
\end{figure*}
\begin{figure*}[!b]
        \includegraphics[width=0.36\columnwidth,height=5.06cm]{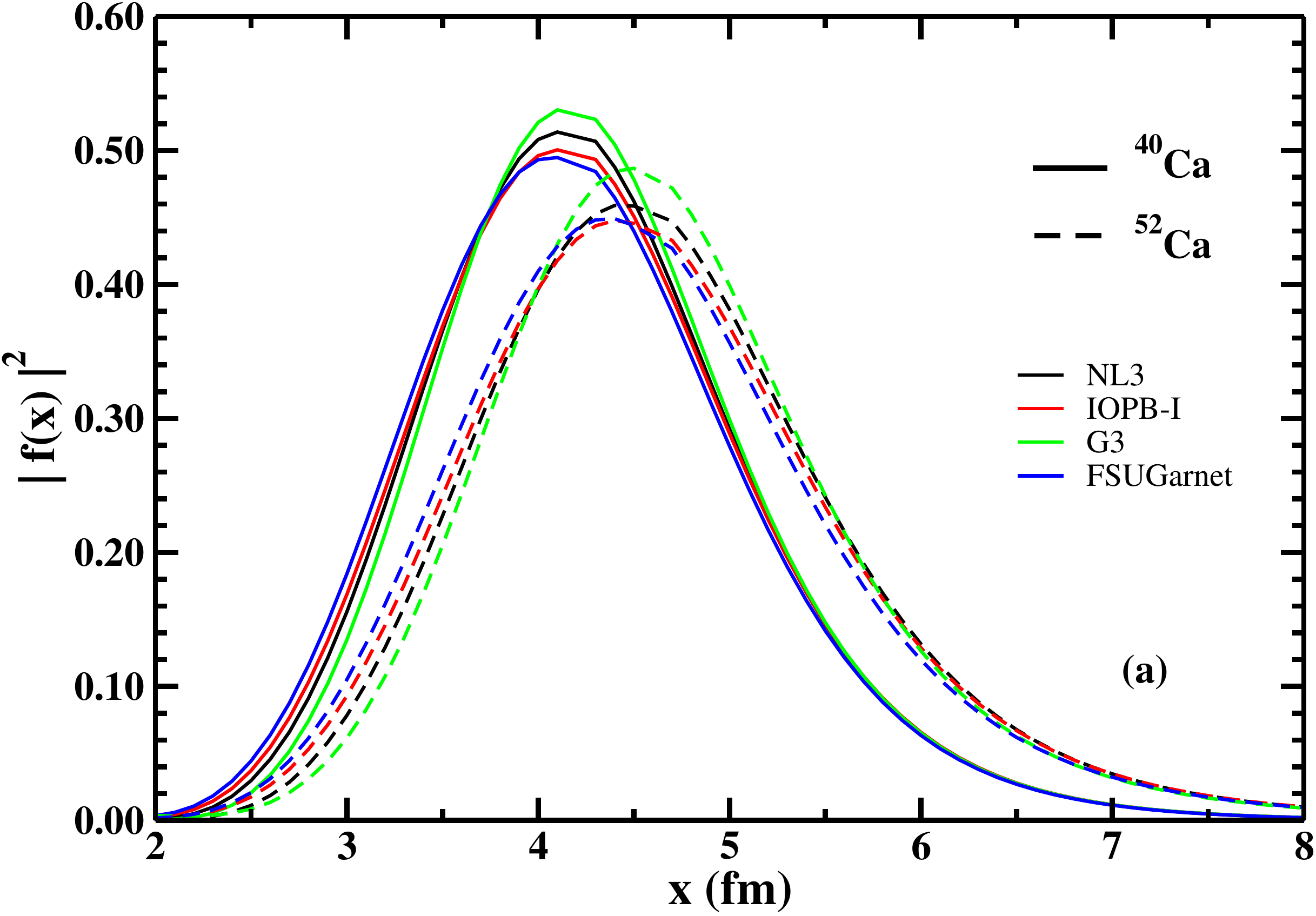}
        \includegraphics[width=0.32\columnwidth,height=5.0cm]{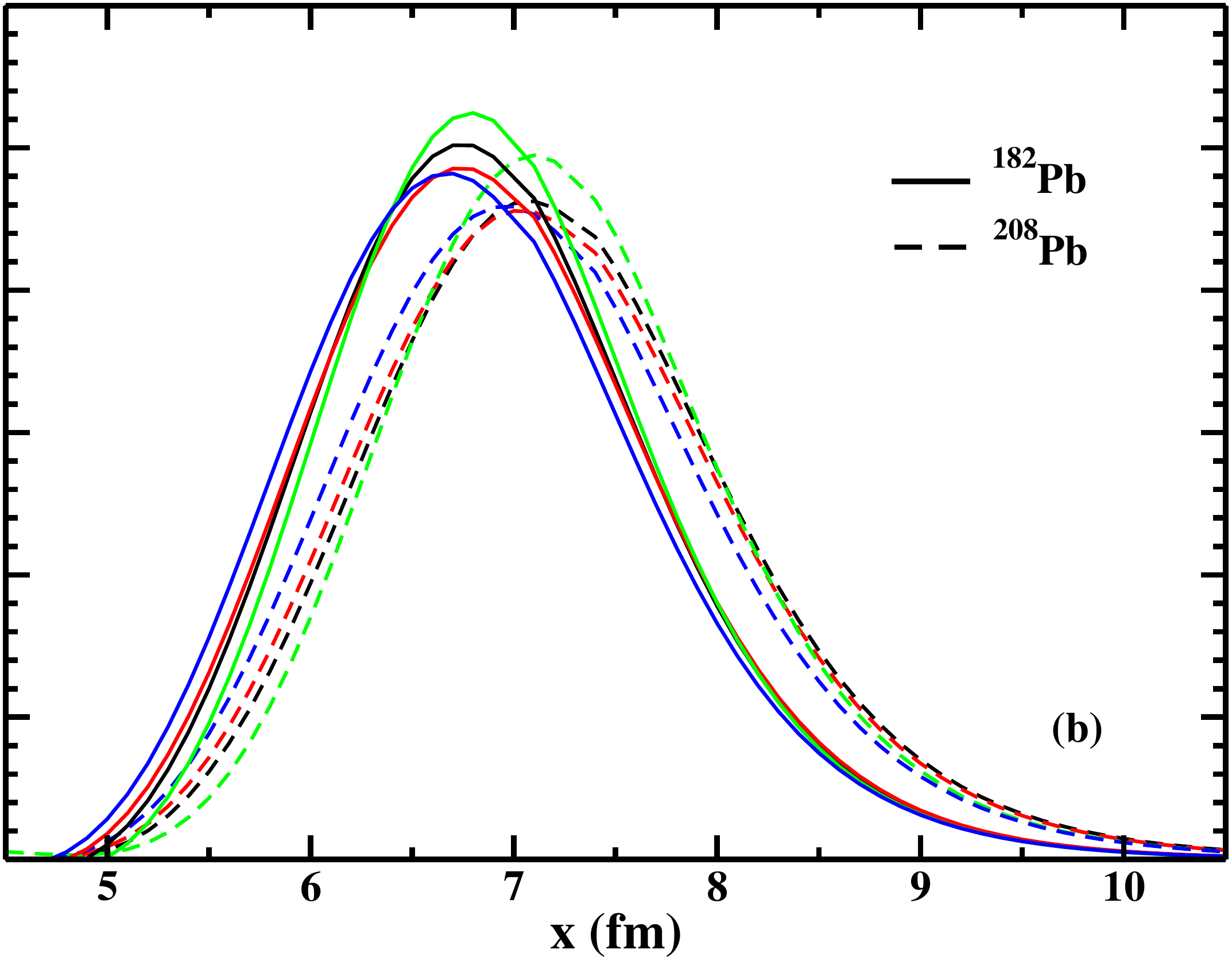}
        \includegraphics[width=0.32\columnwidth,height=5.0cm]{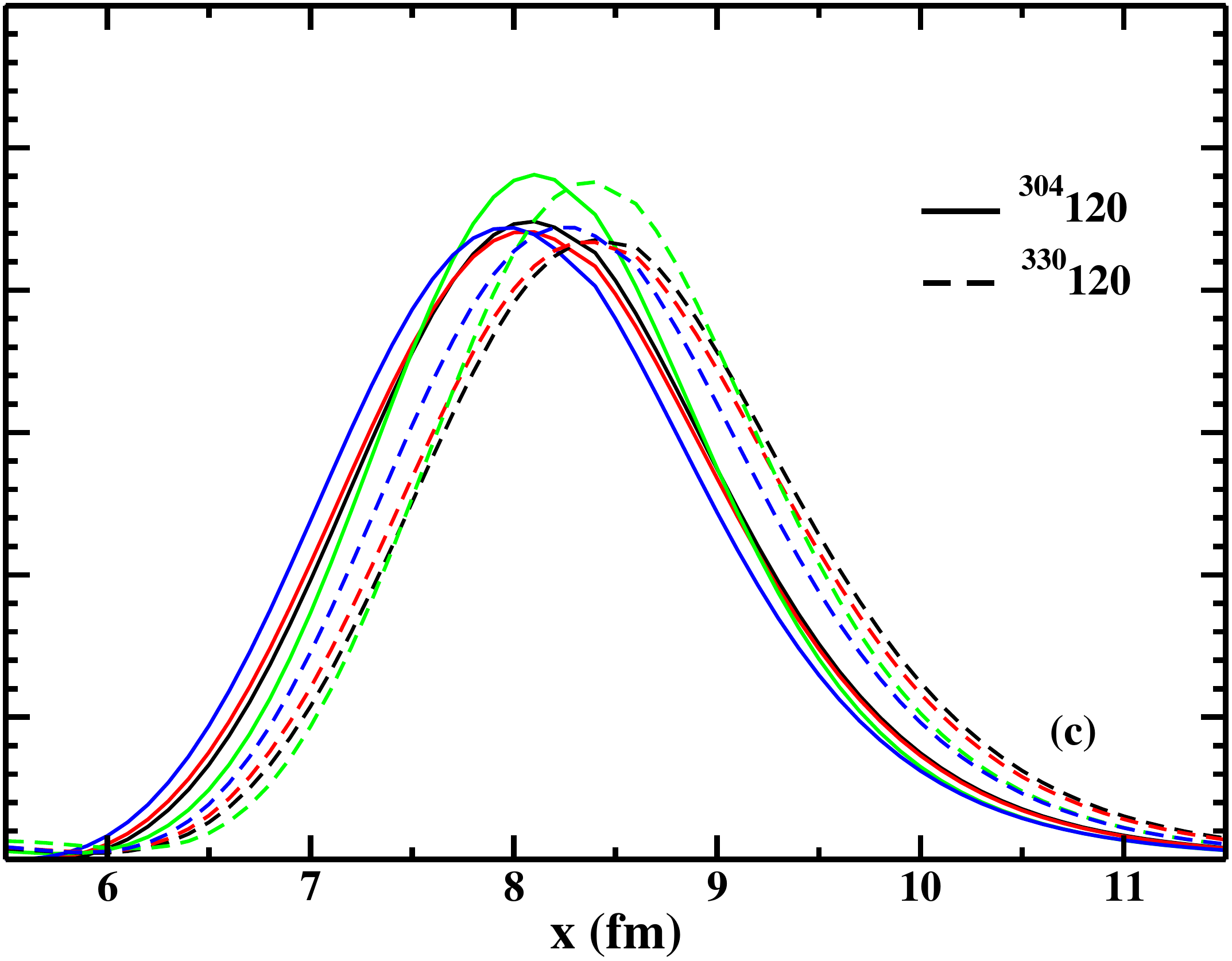}
\caption{(color online) The weight function for (a) $^{40,52}$Ca, (b) $^{182,208}$Pb, and (c) $^{304,330}120$ 
as the representative cases corresponding to  FSUGarnet \cite{chai15}, IOPB-I \cite{iopb1}, G3 \cite{G3}, and 
NL3 \cite{lala97} parameter sets.}
        \label{weight}
\end{figure*}


\begin{figure*}[!b]
        \includegraphics[width=0.36\columnwidth,height=5.05cm]{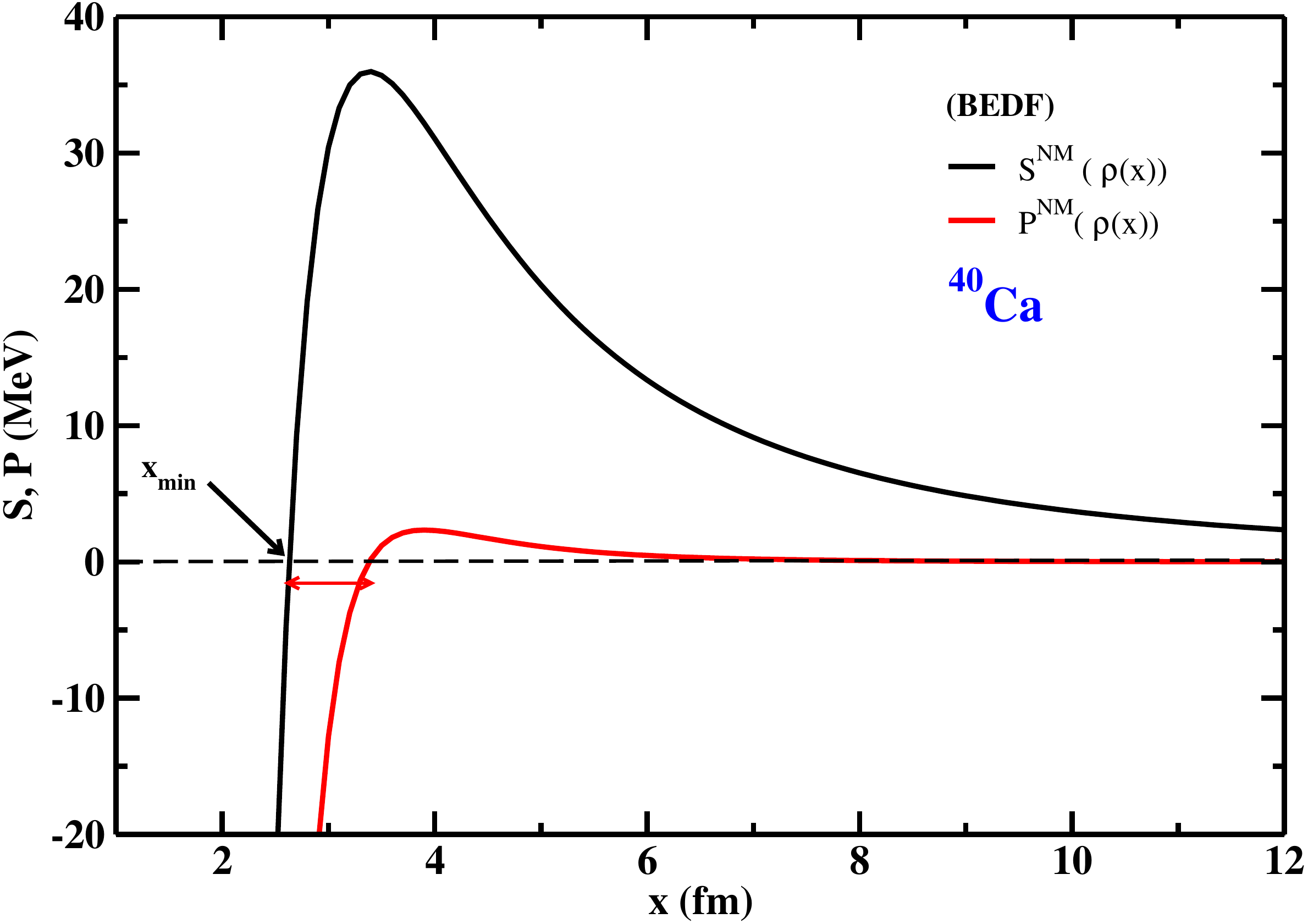}  
        \includegraphics[width=0.32\columnwidth,height=5.0cm]{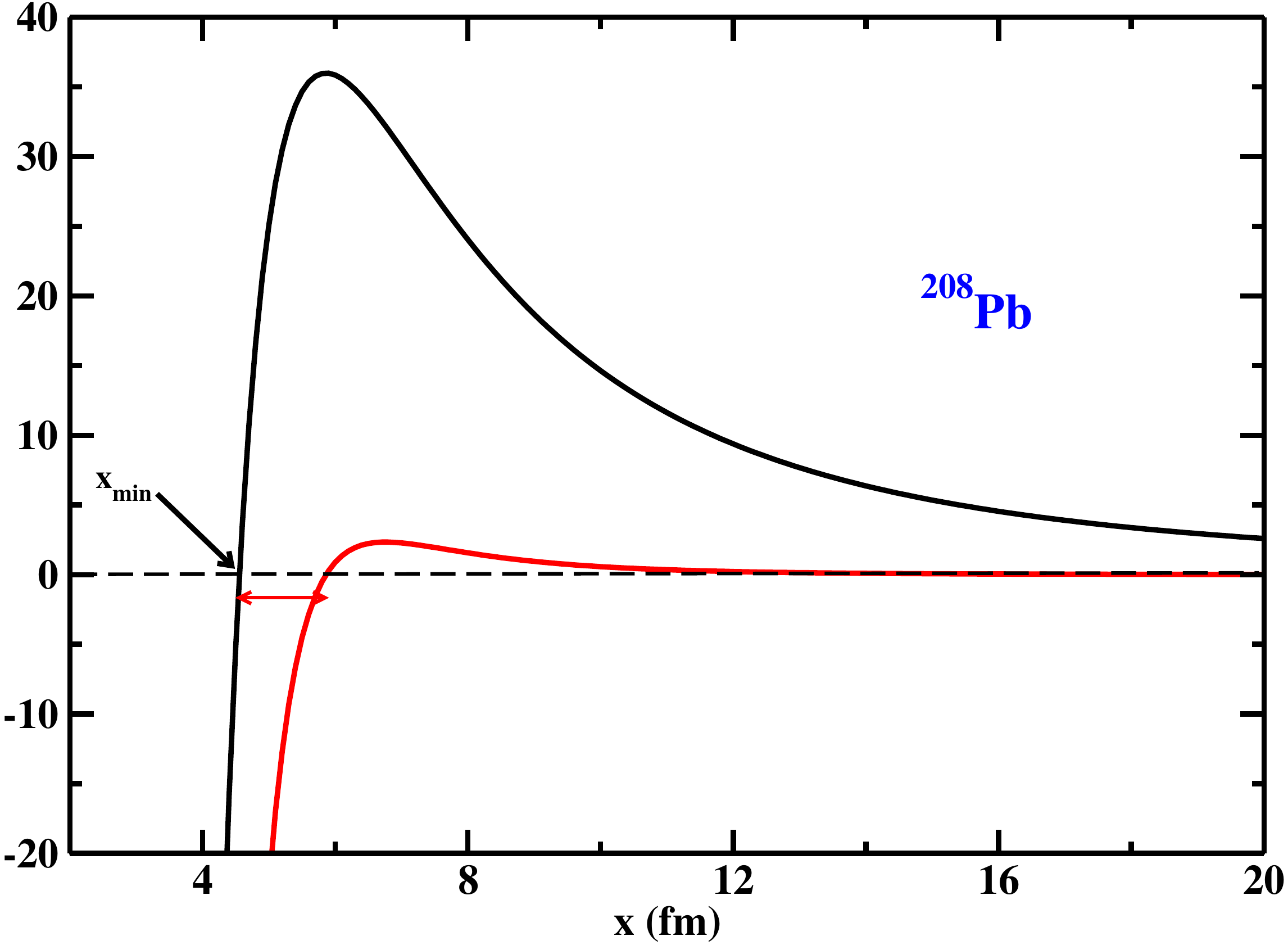}
        \includegraphics[width=0.32\columnwidth,height=5.0cm]{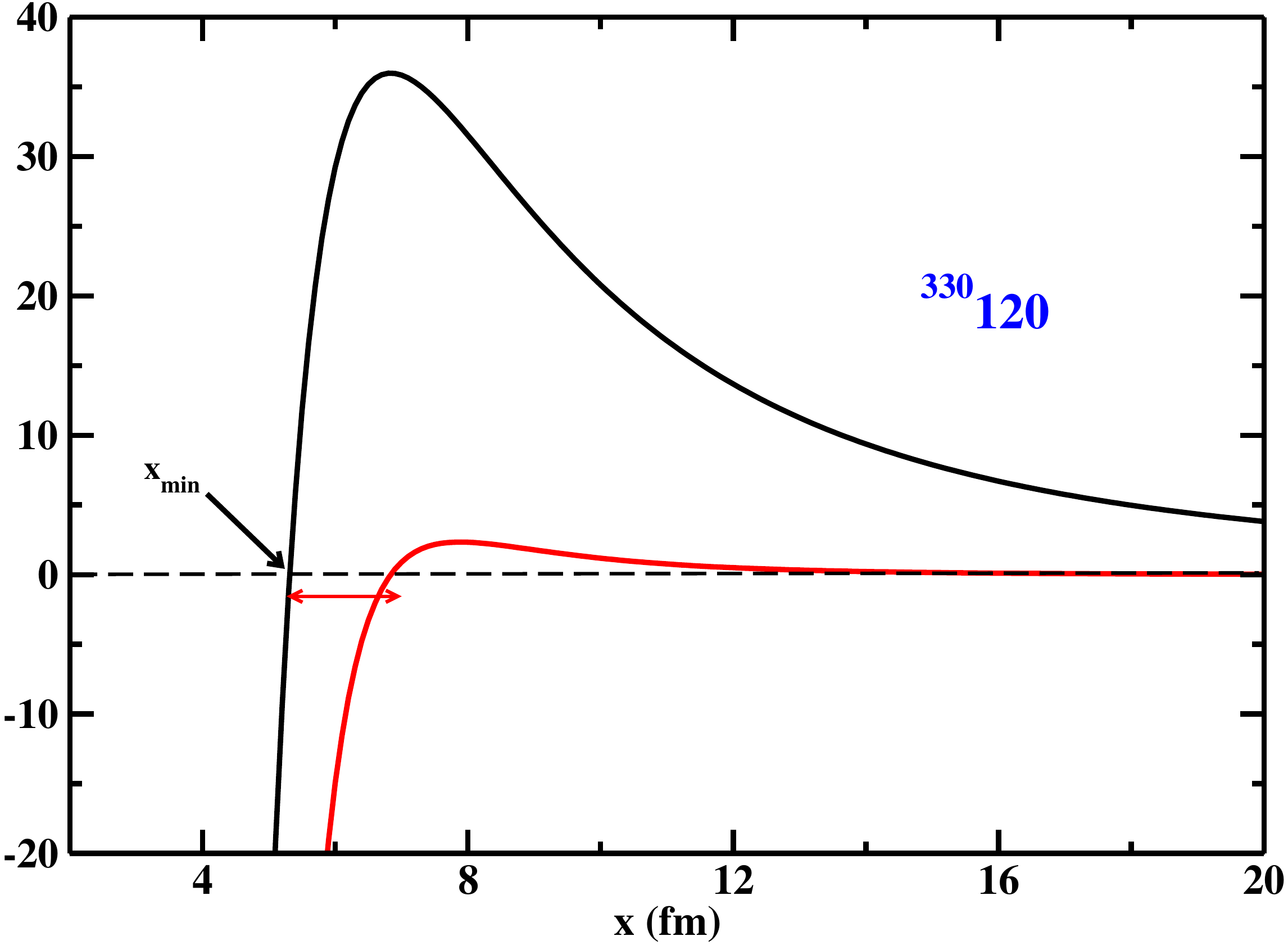}
\caption{(color online) The symmetry energy $S^{NM}$, and the neutron presser $P^{NM}$ of nuclear matter at the Flucton's 
density of (a) Ca, (b) Pb, and (c) Z = 120  within the Brueckner energy density functional (BEDF) as a function 
of local coordinate x. The $x_{min}$ represent the lower limit of the integrations (Eqs. \ref{s0}, 
\ref{p0}, and \ref{k0}). The negative values before the $x_{min}$ points are the unphysical values. The dashed lines 
in the figure mark zero to differentiate between positive and negative values of $S^{NM}$, and $P^{NM}$.}
        \label{nms}
\end{figure*}

\begin{figure}[!b]
        \includegraphics[width=0.50\columnwidth,height=5.50cm]{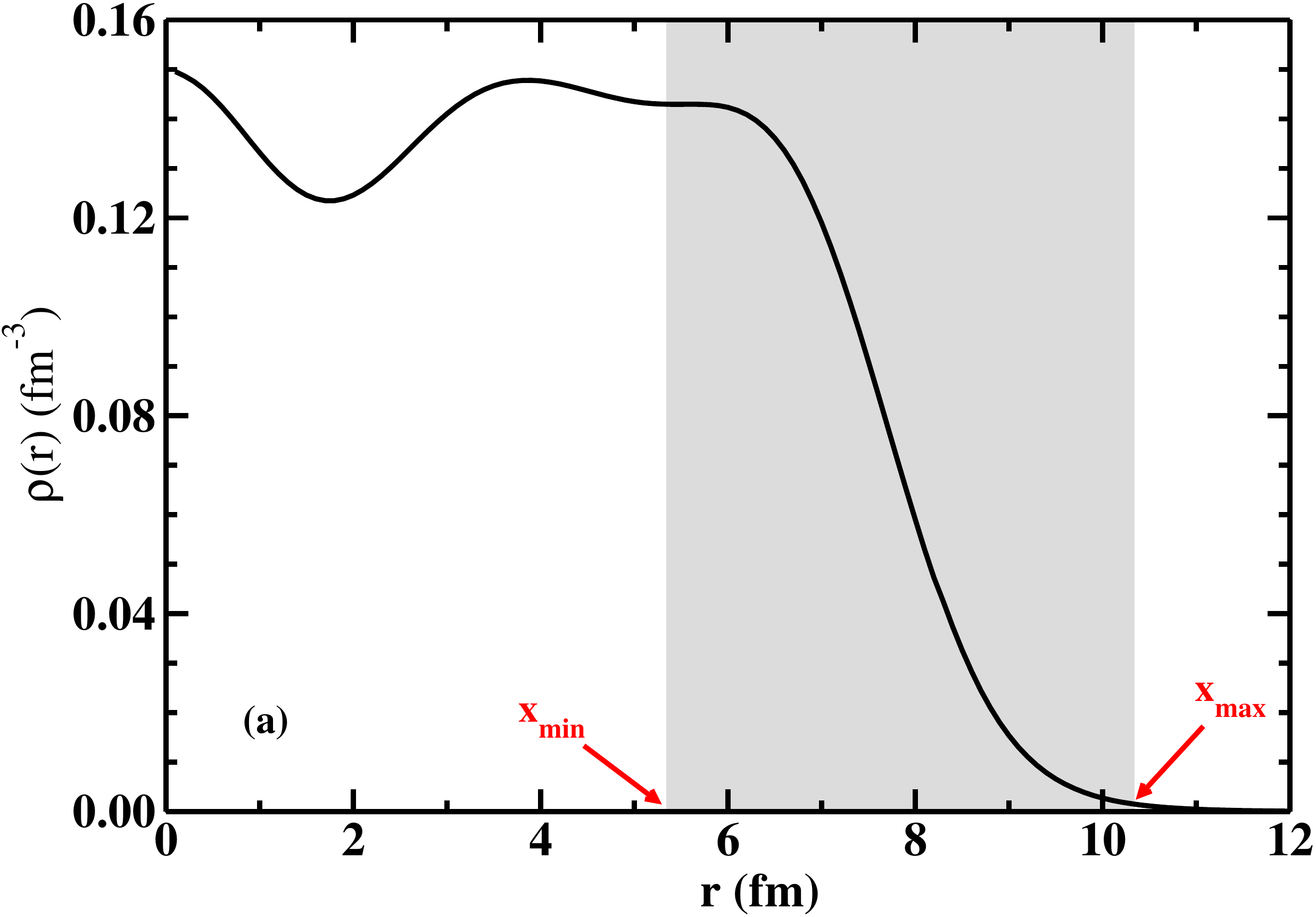}
        \includegraphics[width=0.50\columnwidth]{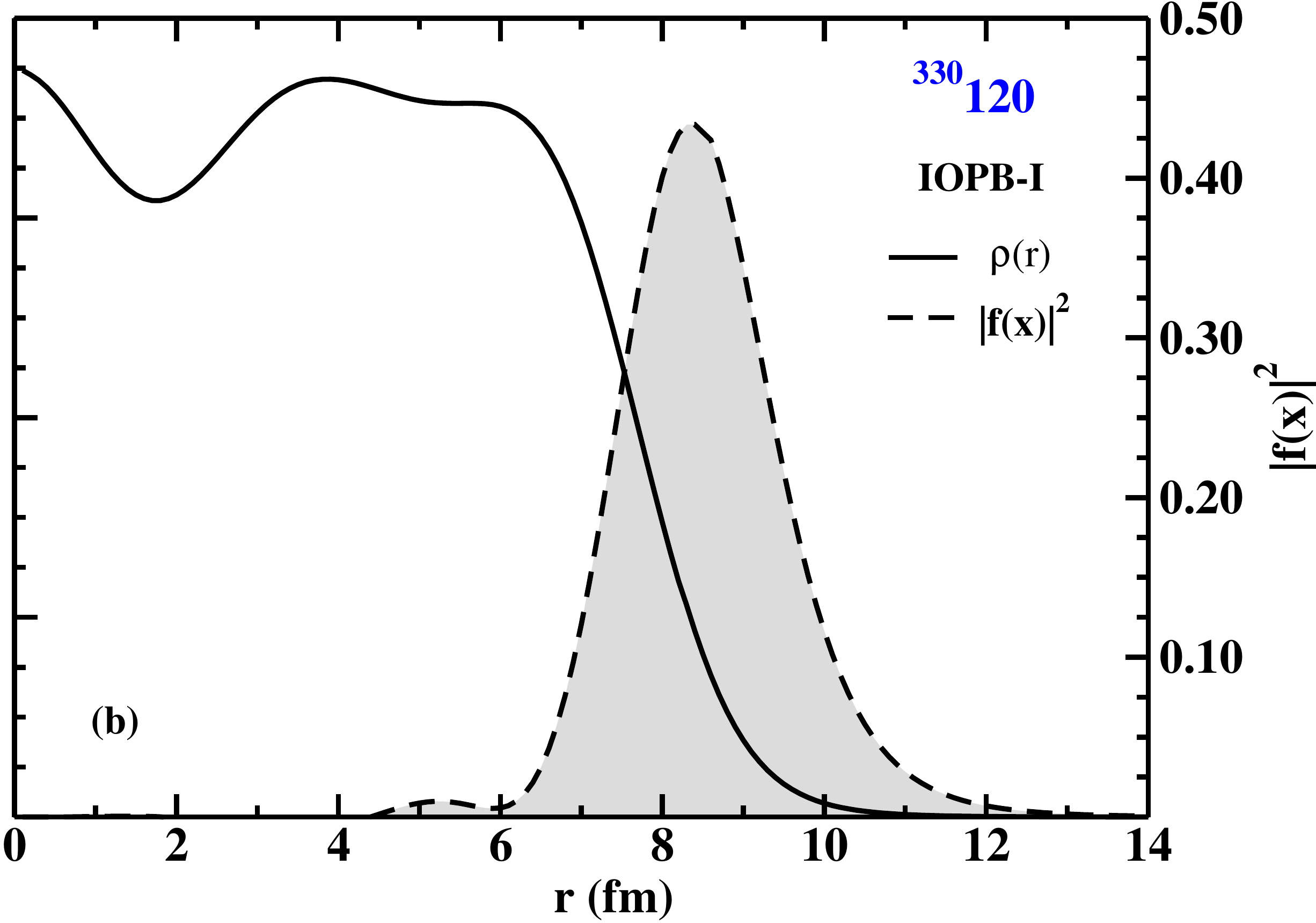}
        \caption{(color online) (a) The density for $^{330}$120 with the limits of integrations (Eqs. \ref{s0}, 
\ref{p0}, and \ref{k0}) $i.e.,$ $x_{min}$ and $x_{max}$, (b) the weight function $\vert f(x) \vert^2$ ($x=r$) 
for $^{330}$120 along with its density,  corresponding to IOPB-I parameter set. 
}
        \label{x1x2}
\end{figure}

\subsection{Ground state properties of the nuclei} {\label {gsp}}
The main aim of this work is to study the effective surface properties like the symmetry energy $S$, 
neutron pressure $P$, and symmetry energy curvature $\Delta K$ for the isotopes (from neutron-deficient 
to the neutron-rich side of the nuclear landscape) of the light, heavy, and superheavy nuclei. Before 
proceeding to the effective surface properties, we have calculated their ground state bulk properties 
within the E-RMF model. Within E-RMF, we have used some of the recent force parameters like FSUGarnet 
\cite{chai15}, IOPB-I \cite{iopb1}, and G3 \cite{G3}. The FSUGarnet \cite{chai15}, IOPB-I \cite{iopb1}, 
and G3 \cite{G3} parameters have the advantage that their EoSs are softer compared to the NL3 parameter. 
The motive behind choosing these parameter sets has been illustrated in Ref. \cite{aqjpg}. 
Table \ref{table1} contains the values of these parameter sets with the key EoS parameters for the nuclear 
matter at saturation (lower panel). From the table, we notice that the symmetry energy ($J$) and its 
coefficients $L$, $K_{sym}$, and Q$_{sym}$ are consistent with the allowed empirical and/or experimental 
ranges for three force parameters, namely FSUGarnet, G3, IOPB-I. The allowed empirical and/or experimental 
ranges along with the non-relativistic and relativistic constraints for a large number of force parameter sets 
can be found in Refs. \cite{dutra12,dutra14}. In case of NL3 force parameter, these values are overestimated 
to the allowed constraint ranges. These overestimations are due to the well-known stiffer nature of EoS. Furthermore, 
one can find that these ranges get broad with respect to density as well as higher order derivatives (i.e, 
the derived quantities from symmetry energy), which are the accepted behavior of mean field models. 
The binding energy per particle (B/A), charge radius ($R_{ch}$), and neutron skin-thickness 
($\Delta r=R_n - R_p $) of some of the double magic nuclei for FSUGarnet \cite{chai15}, IOPB-I \cite{iopb1}, 
G3, and NL3 \cite{lala97} parameter sets are listed in Table \ref{table3} with the available experimental 
data \cite{audi12,angeli13}. The calculated properties of finite nuclei corresponding to all chosen parameter 
sets are in good agreement with each other. These results are comparable to the available experimental data. 
On inspecting the table, it is found that in some cases, the binding energy corresponding to the IOPB-I 
force parameter set overestimates the experimental data. 

\subsection{Densities and weight functions for the nuclei}
\label{denwef}
It is important to note that the measurement of neutron density distribution in a nucleus is difficult 
task because of its neutral nature. As a result, the determination of neutron distribution radius R$_n$ has 
been poorly done. On the other hand, the proton distribution radius R$_p$ has been measured with a high accuracy. 
To solve this problem the recently proposed parity violation experiments PREX-II at JLab \cite{prex1} and the Bates 
Laboratory at MIT were done with polarized beams and targets \cite{bate} which have given better results for the neutron 
distribution \cite{prex1,bate}. The anti-proton experiment at CERN gives the neutron skin-thickness for 26 stable nuclei 
starting from $^{40}$Ca to $^{238}$U \cite{jast}. The recently reported G3 and IOPB-I force parameters in the framework of 
the E-RMF formalism reproduce the neutron skin-thickness $\Delta r$ quite well \cite{iopb1}. This gives us 
confidence that although the measurement of neutron distribution inside the nucleus is not as general as the proton 
even then our chosen forces should be capable enough to reproduce the proton and neutron distribution and 
hence the total density distribution of a nucleus.

The total densities of the nuclei, calculated within the spherically symmetric E-RMF formalism corresponding to the 
NL3, IOPB-I, G3, and FSUGarnet parameter sets, are shown in Fig. \ref{dens}. The color code is represented 
in the legends. The bold lines represent the total densities for neutron-deficient nuclei while dashed lines represent 
the densities for neutron-rich isotopes. The panels (a), (b), 
and (c) of the figure show the densities of $^{40,52}$Ca, $^{182,208}$Pb, and $^{304,330}$120, respectively,
as the representative cases. It can be noticed from the figure that the central part of the density is larger 
for lighter isotopes than those of heavier isotopes for particular nuclei. On the other hand, the surface 
densities are enhanced a bit as a function of radius for heavier isotopes than lighter. 

The calculated densities from the E-RMF model are further used in Eq. \ref{wfn} to obtain the weight 
functions for the corresponding nuclei. The weight functions for the nuclei $^{40,52}$Ca, $^{182,208}$Pb, 
and $^{304,330}$120 as the representative cases are shown in the panels (a), (b), and (c) of Fig. 
\ref{weight}, respectively. From the figure, one can notice that the trend of the density profile (Fig. \ref{dens}) is reflected 
in weight functions. In other words, the lower value of the central 
density gives the lesser height of the weight function for an isotope of the particular nucleus. Further, 
it can be noticed in the figure that the maxima of the weight functions shift towards the right (larger $r$) 
with the size of a nucleus increases. The G3 parameter set predicts the larger weight function for all the 
nuclei while the lower one corresponds to the FSUGarnet parameter set. The symmetry energy, neutron pressure 
and symmetry energy curvature of nuclear matter at local coordinate are folded with the calculated weight 
function of a nucleus which results in the corresponding effective surface properties of the finite nucleus. 
It would be worth illustrating the point that why these quantities are termed as the surface properties 
and how to find the limits of integration (Eqs. \ref{s0}-\ref{k0}).

In principle, the limits of integration in Eqs. \ref{s0}-\ref{k0} are set from $0$ to $\infty$. But, the 
symmetry energy of infinite symmetric nuclear matter within the Brueckner energy density functional method has 
some negative values (unphysical points) in certain regions. In order to avoid the unphysical points of the 
symmetry energy of nuclear matter, the limits of integration $x_{min}$ and $x_{max}$ are put other than what 
mentioned above. In general, $x_{min}$ and $x_{max}$ are the points where the symmetry energy of nuclear 
matter changes from negative to positive and from positive to negative, respectively \cite{gai12}. For the 
better understanding of the concept of finding $x_{min}$ and $x_{max}$, we present the symmetry energy of 
nuclear matter within Brueckner energy density functional $S^{NM}(x)$ (used in Eq. \ref{s0}) for $^{40}Ca$, 
$^{208}Pb$, and $^{330}120$ in the panels (a), (b), and (c) of Fig. \ref{nms}. The nature of the curves for 
the symmetry energy of nuclear matter is the same for the nuclei shown in the figure with almost the same maximum 
value. However, the curves shift towards the right (larger values of $r$) with nuclei having a large mass 
number. It can easily be noticed from the figure that at $x=2.5$fm, $x=4.3$fm, and $x=5.2$ fm, the 
$S^{NM}(x)$ of nuclear matter changes from negative to positive at Flucton density in $^{40}Ca$, 
$^{208}Pb$, and $^{330}120$, respectively. Thus, these points are considered as $x_{min}$ for the respective 
nuclei. While no point at large $x$ seems to be such that where the $S^{NM}(x)$ changes from positive to 
negative (in this case). Rather, $S^{NM}(x)$ tends to zero at large values of $x$. Thus, the value of 
$x_{max}$ can not be fixed in this way. On the other hand, the densities of the nuclei become almost zero 
at $r = 6.0$fm, $r = 9.4$fm, $r = 10.5$fm for $^{40}Ca$, $^{208}Pb$, and $^{330}120$, respectively. 
Therefore, these points are considered as the $x_{max}$ (upper limit of integrations (Eqs. \ref{s0} - 
\ref{k0})). Figure \ref{x1x2} represents the density of $^{330}$120 with the IOPB-I parameter set as the 
representative case, showing the limits of integration $x_{min}$ and $x_{max}$. The limits are used in 
Eqs. (\ref{s0}-\ref{k0}) to find the symmetry energy, neutron pressure, and symmetry energy curvature, 
respectively. It can be noticed from Fig. \ref{x1x2} that the values of limit do not have any central part 
of the density and lie in the surface region. Hence, the quantities S, P, and $\Delta K$ are known as  
surface properties. For further illustrating the concept of referring these quantities as the surface 
properties, the panel (b) of Figure \ref{x1x2} is presented, here, showing the density and weight function 
altogether. It has already been mentioned that the properties of infinite nuclear matter are folded with the 
weight function to obtain the corresponding quantities of finite nuclei. The significant values of weight 
functions (its peak value) lie in the range which corresponds to the surface part of the density. This is 
also one of the reasons to call these quantities as the surface properties. 

\begin{figure*}[!b]
        \includegraphics[width=0.5\columnwidth]{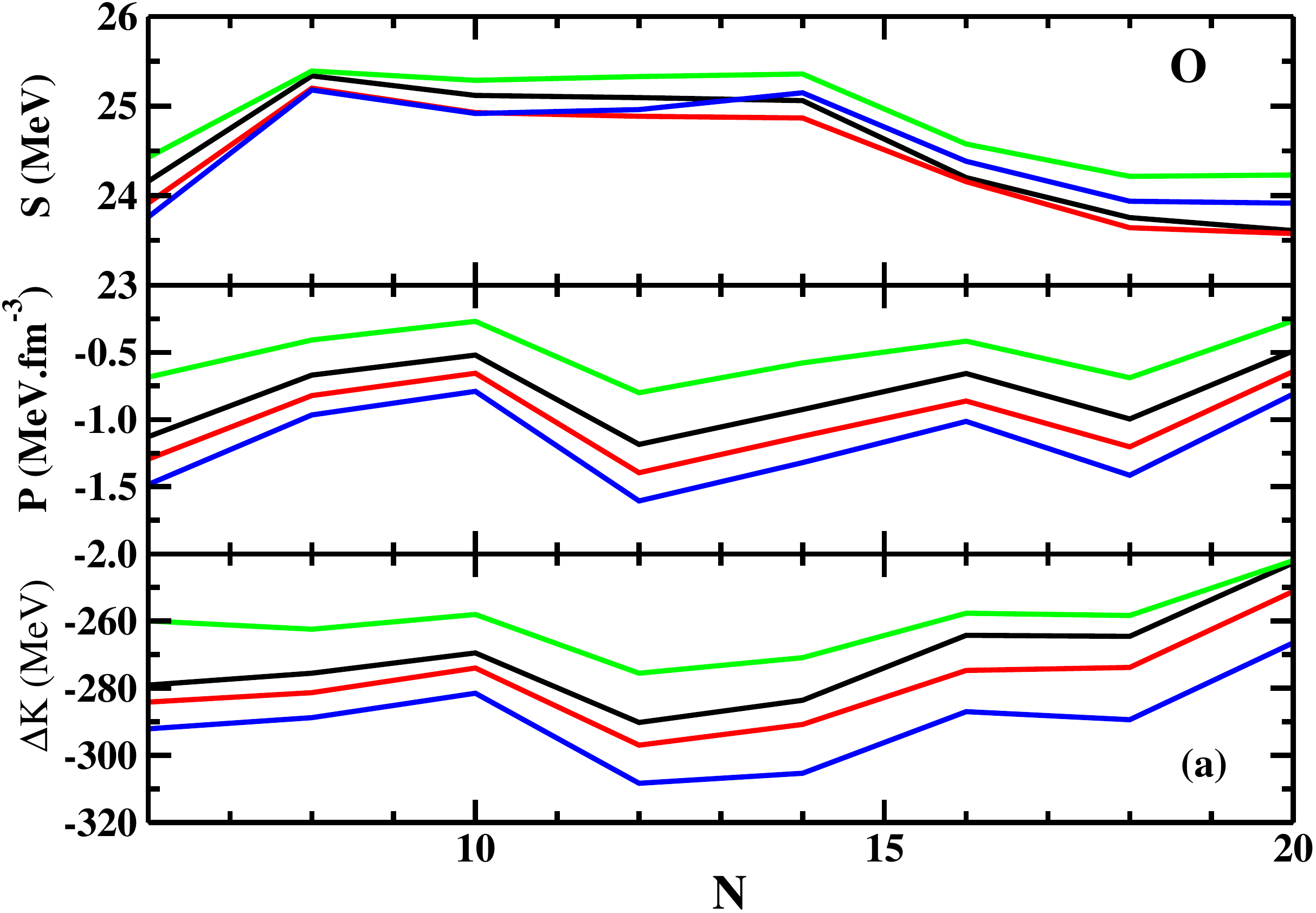}
        \includegraphics[width=0.5\columnwidth]{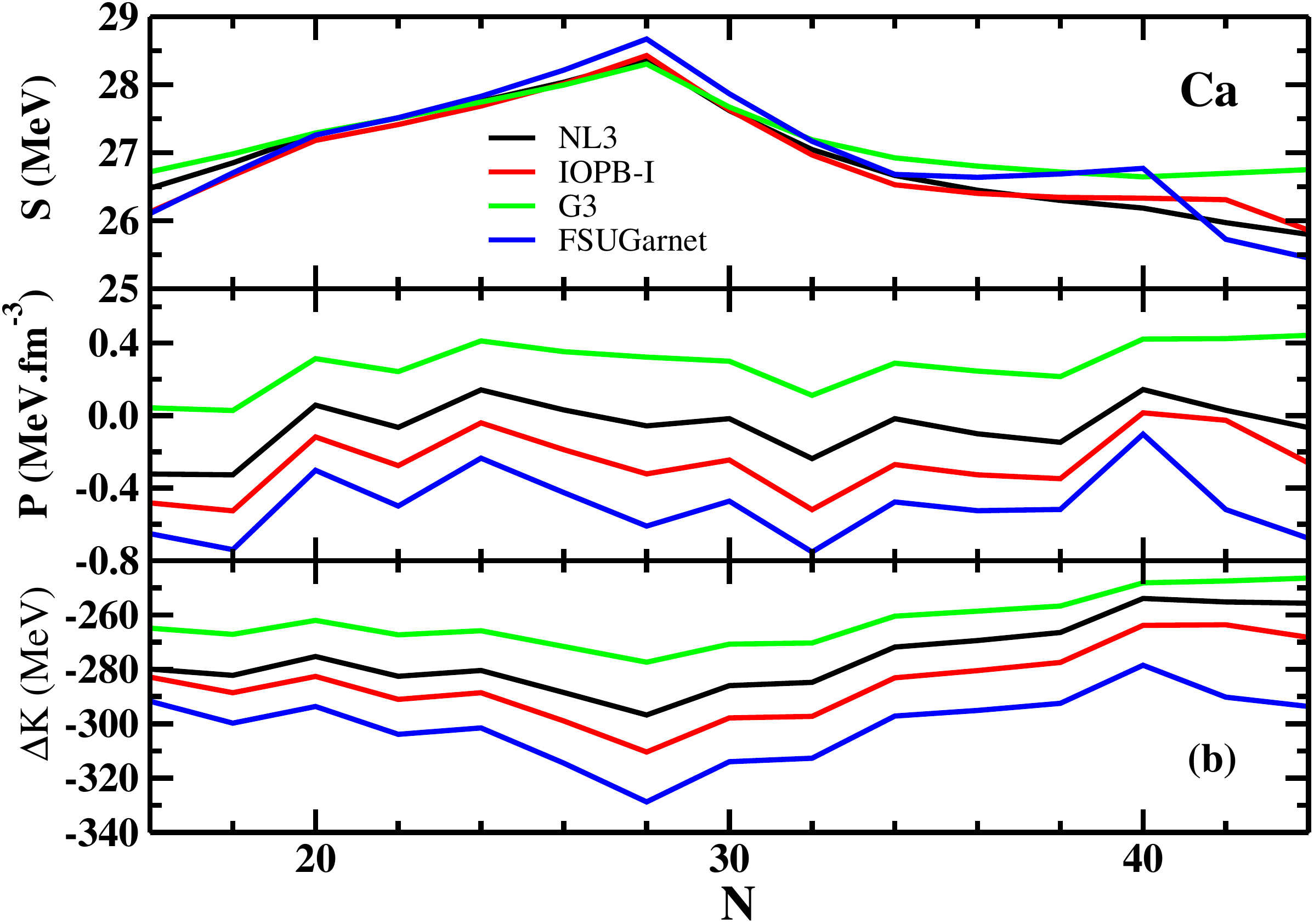}
        \includegraphics[width=0.5\columnwidth]{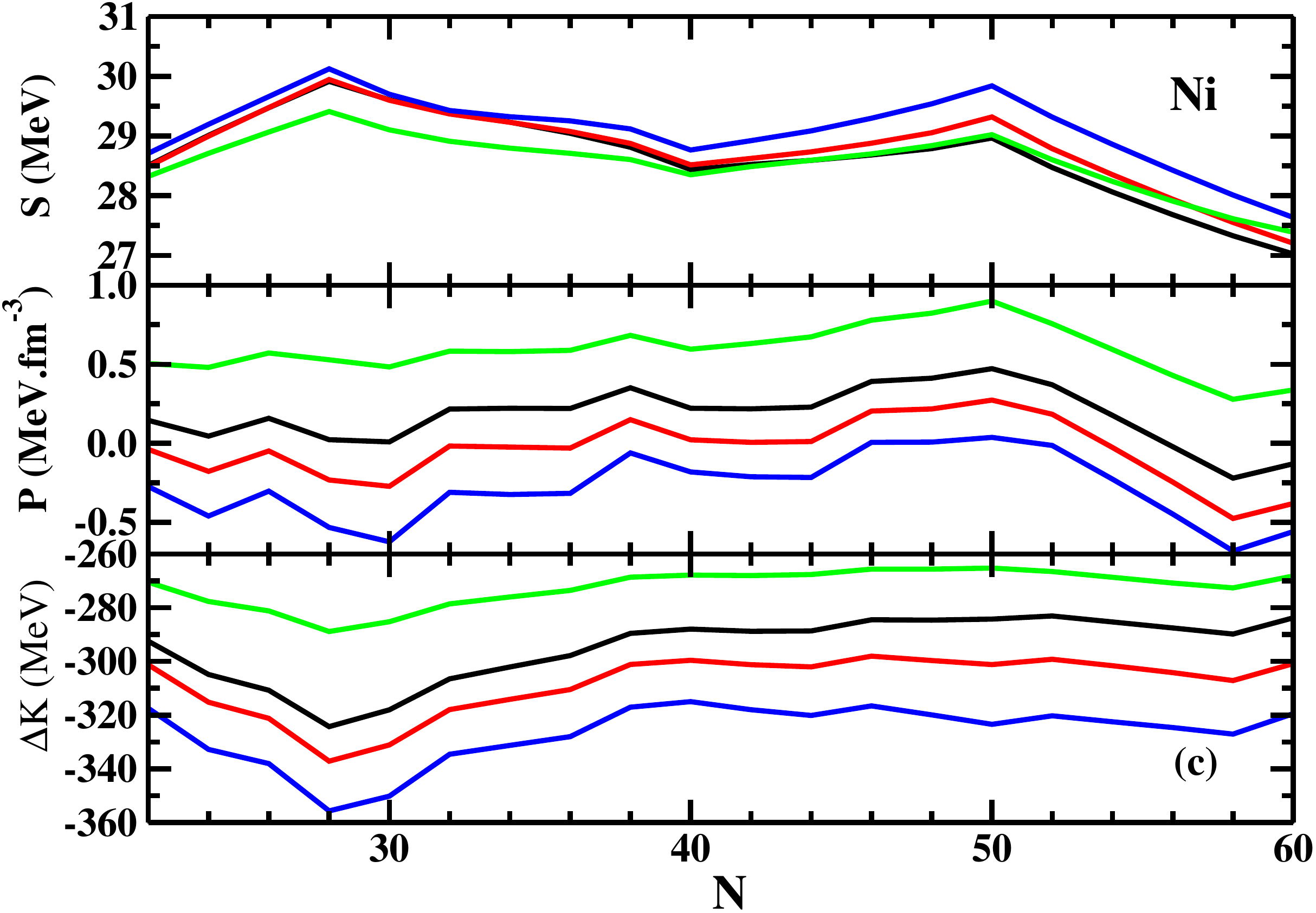}
        \includegraphics[width=0.5\columnwidth]{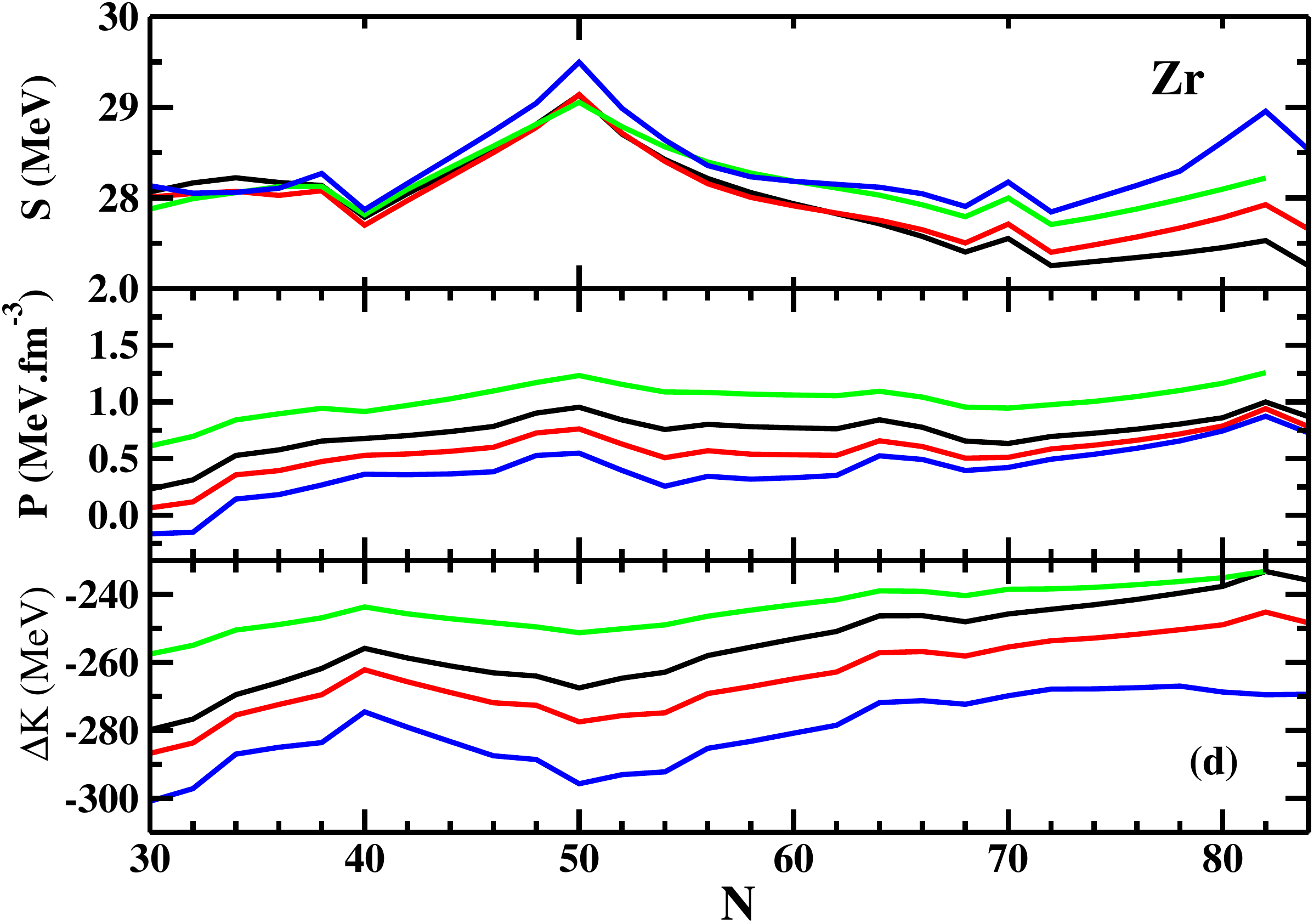}
        \includegraphics[width=0.5\columnwidth]{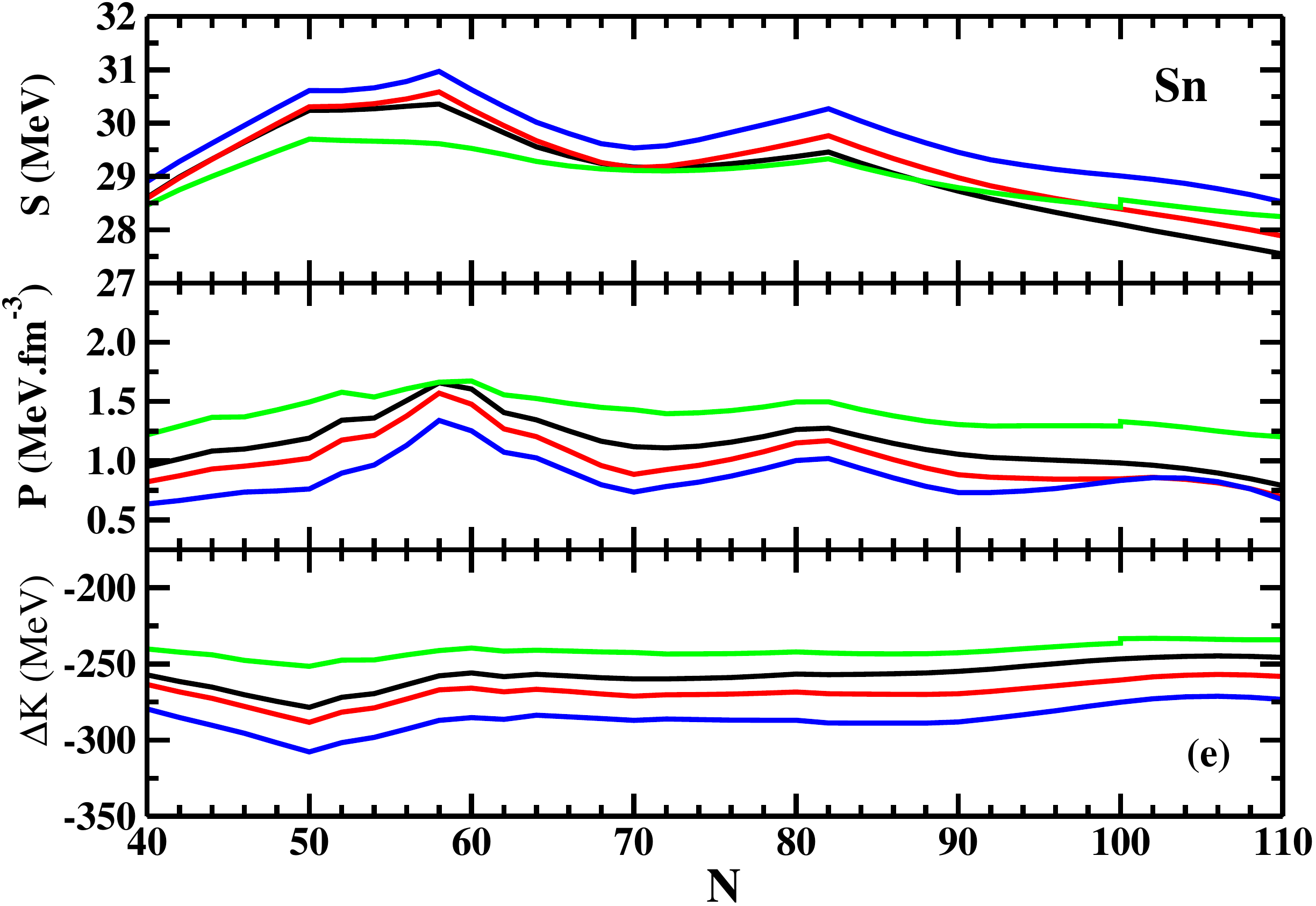}
        \includegraphics[width=0.5\columnwidth]{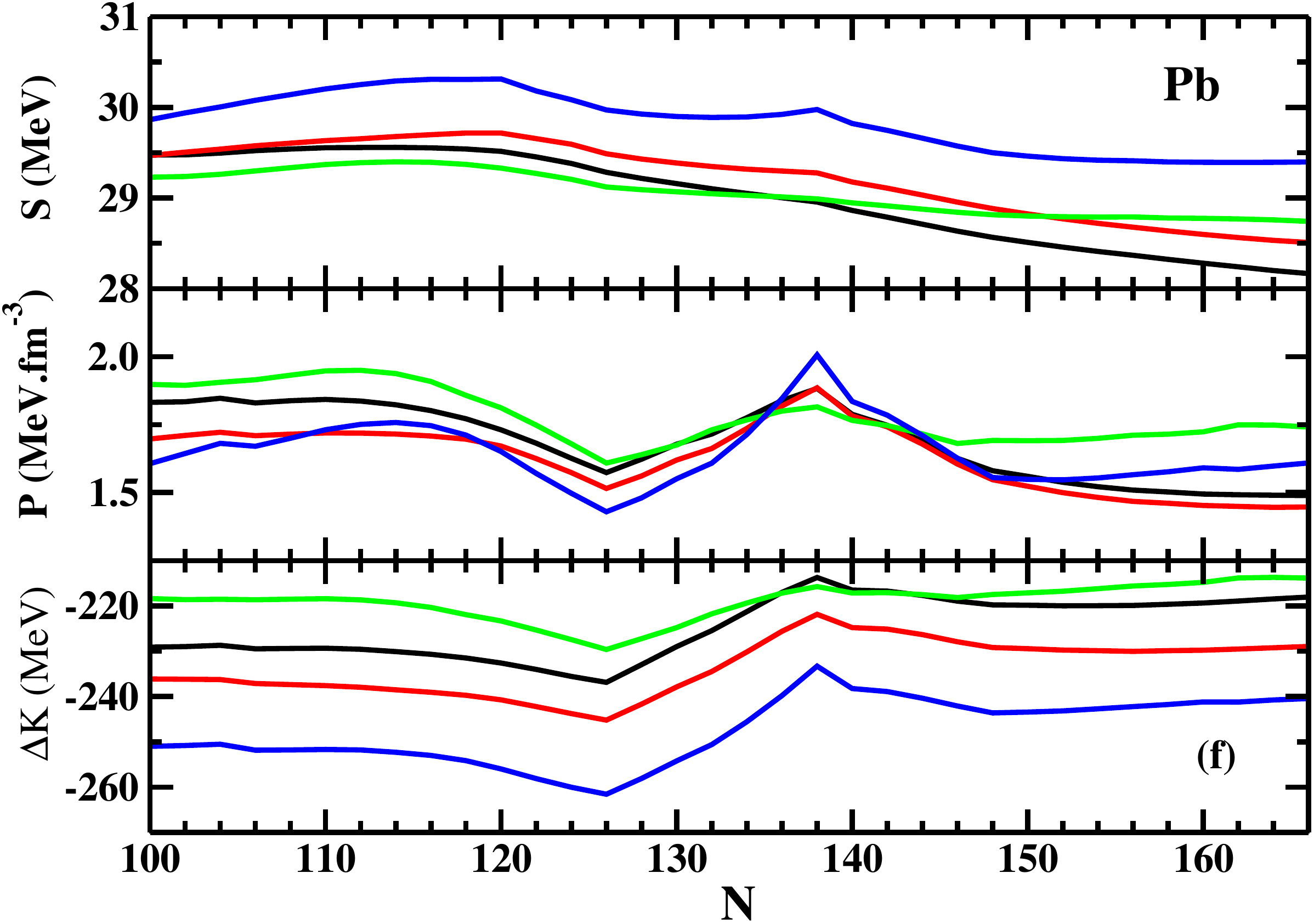}
        \caption{(color online) The symmetry energy ($S$), pressure ($P$), and symmetry energy curvature 
($\Delta K$) for the isotopic series of O, Ca, Ni, Zr, Sn, and Pb nuclei corresponding to NL3, IOPB-I, 
G3, and FSUGarnet parameters sets.}
        \label{spkn}
\end{figure*}

\begin{figure}[!b]
	\centering
        \includegraphics[width=0.5\columnwidth]{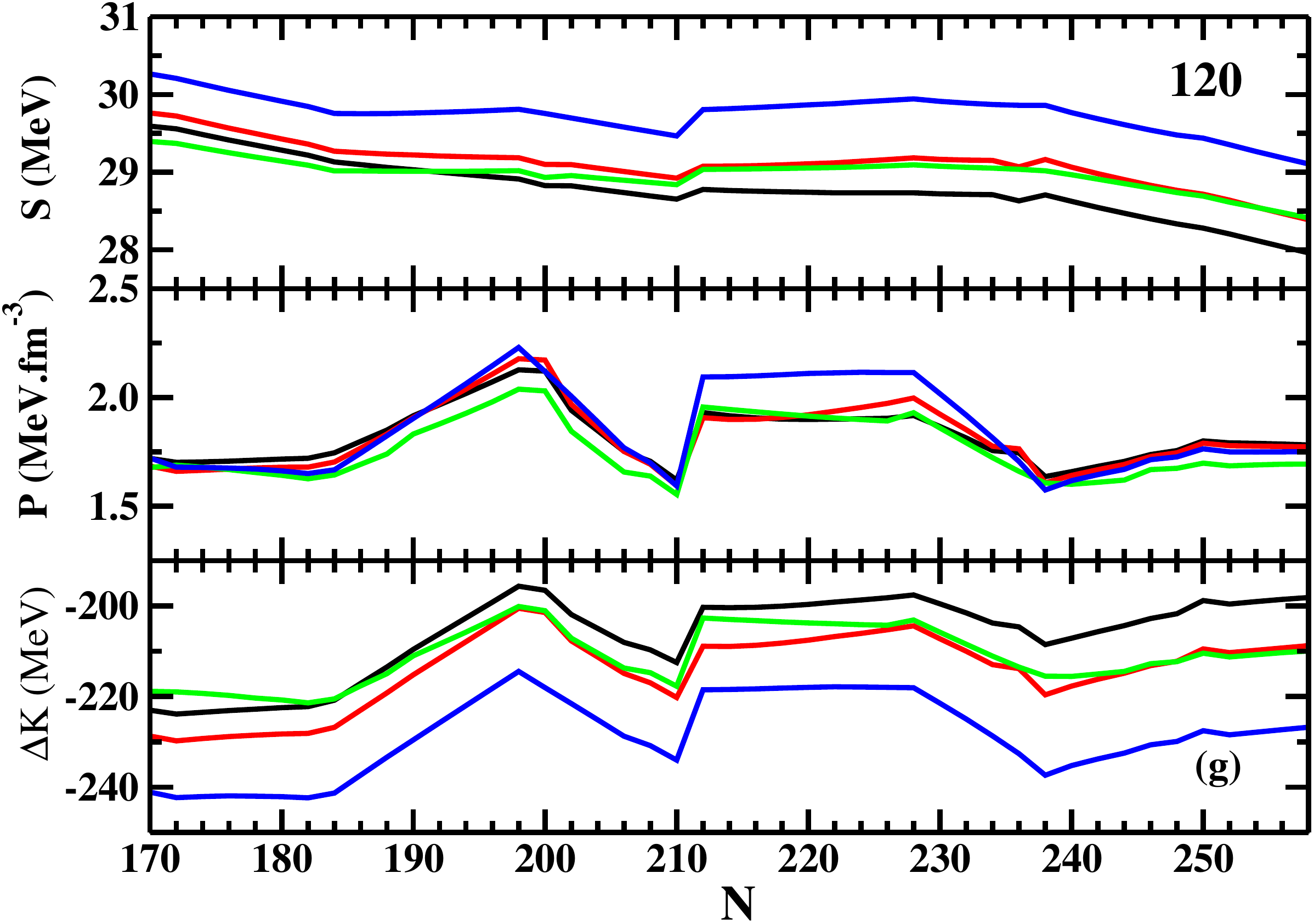}
        \caption{(color online) The description is same as in Fig. \ref{spkn} but for Z = 120 nuclei. }
        \label{120spkn}
\end{figure}

\begin{figure*}[!b]
        \includegraphics[width=0.5\columnwidth]{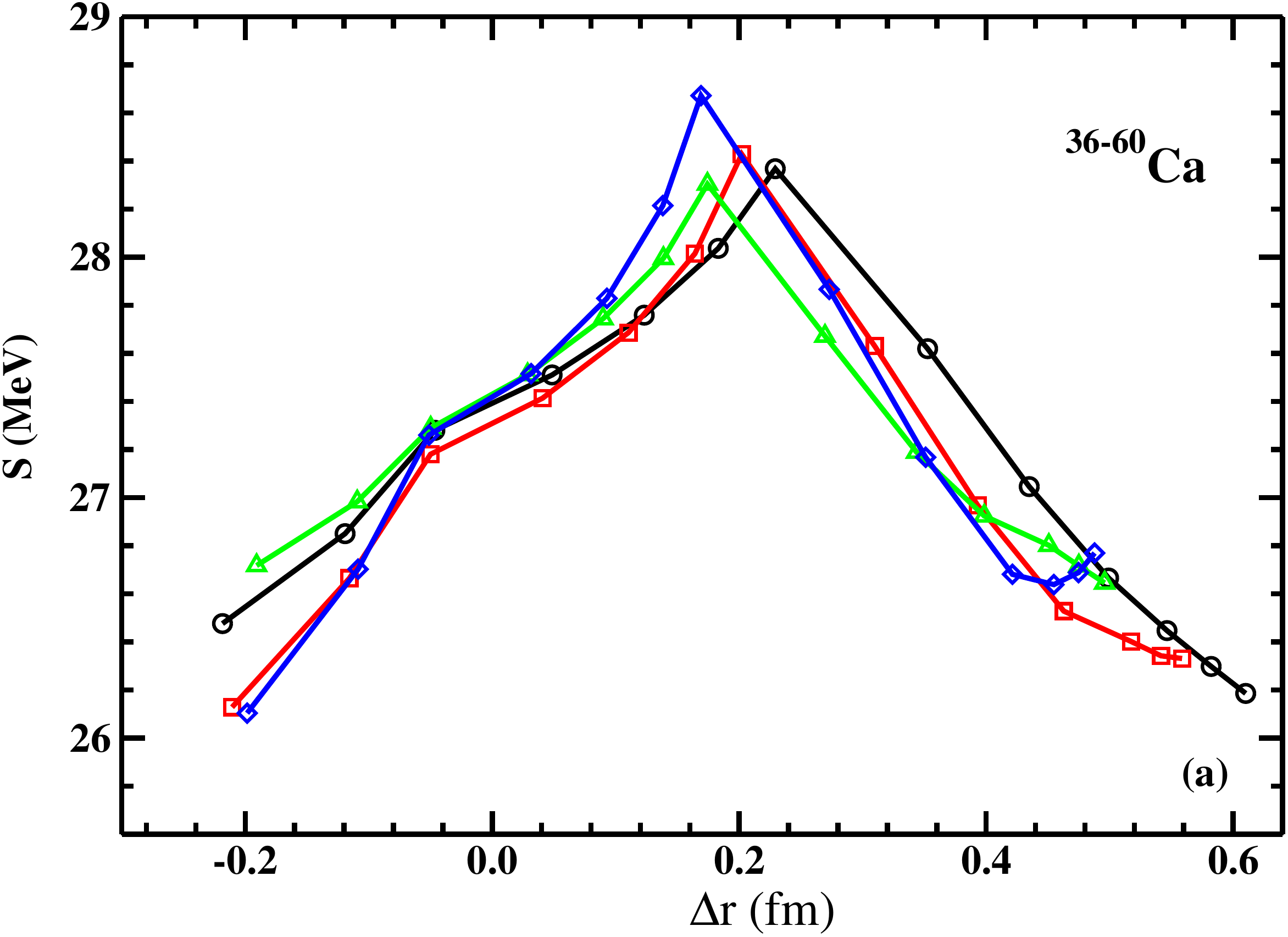}
        \includegraphics[width=0.5\columnwidth]{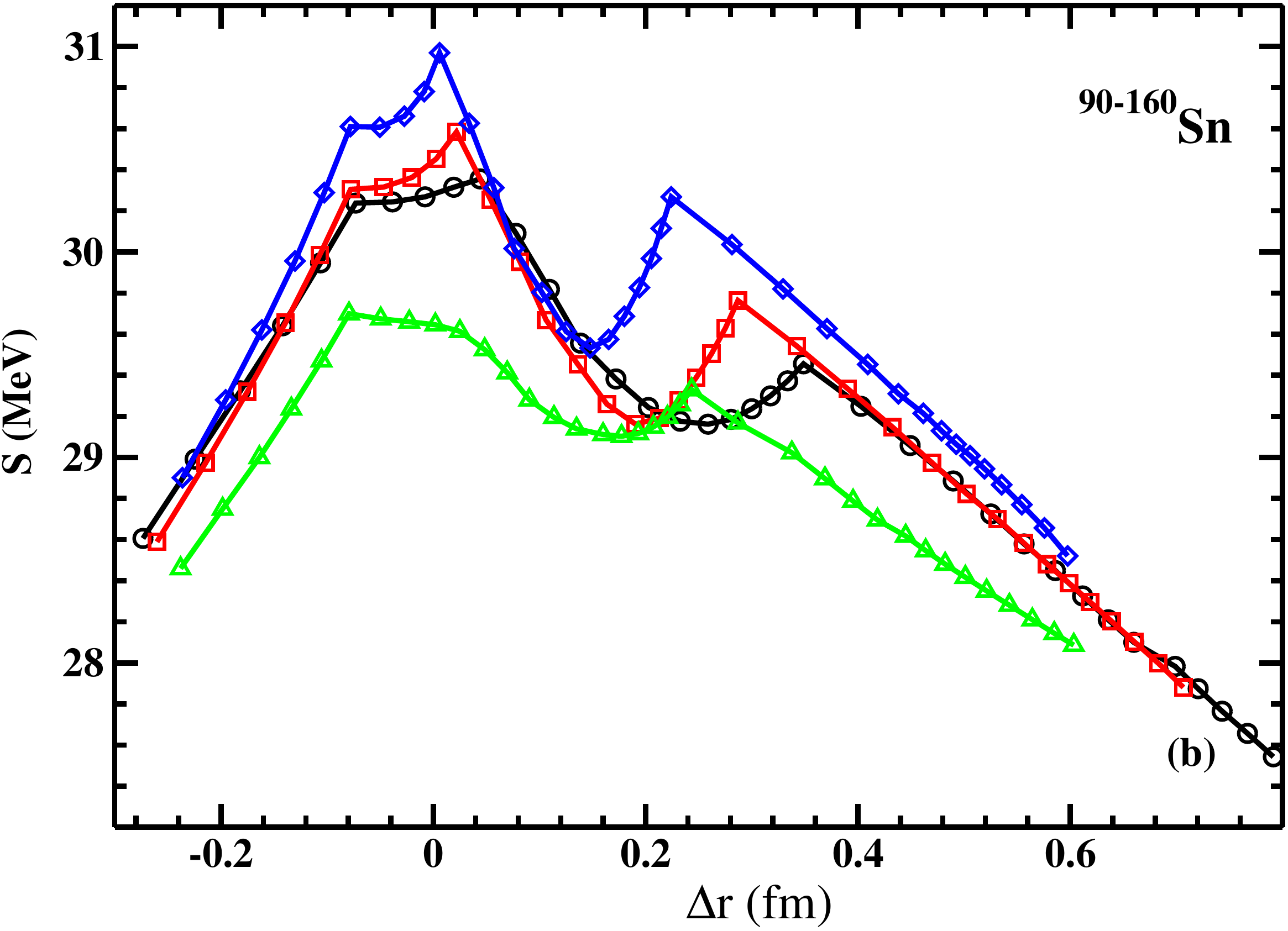}
        \includegraphics[width=0.5\columnwidth]{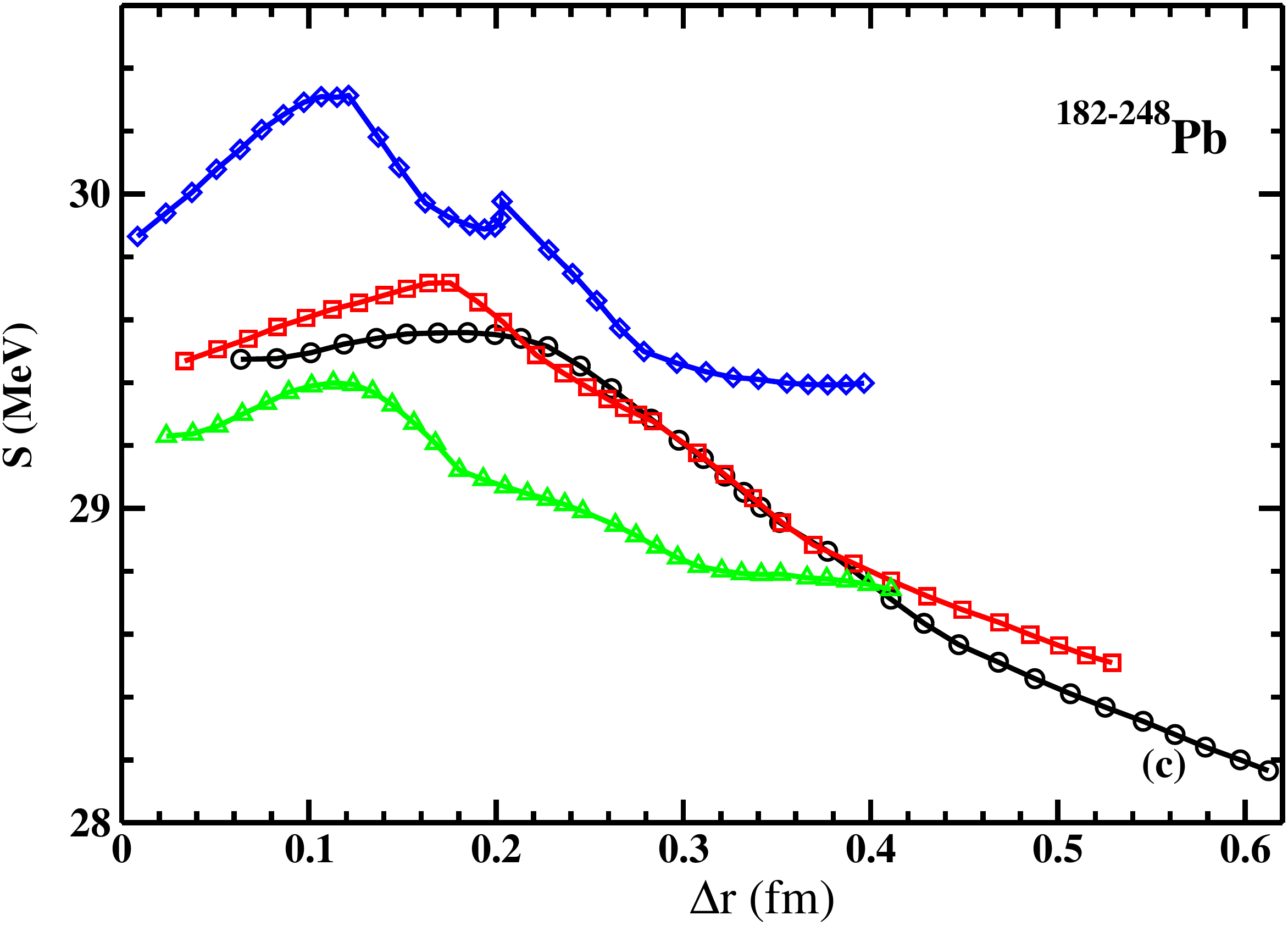}
        \includegraphics[width=0.5\columnwidth]{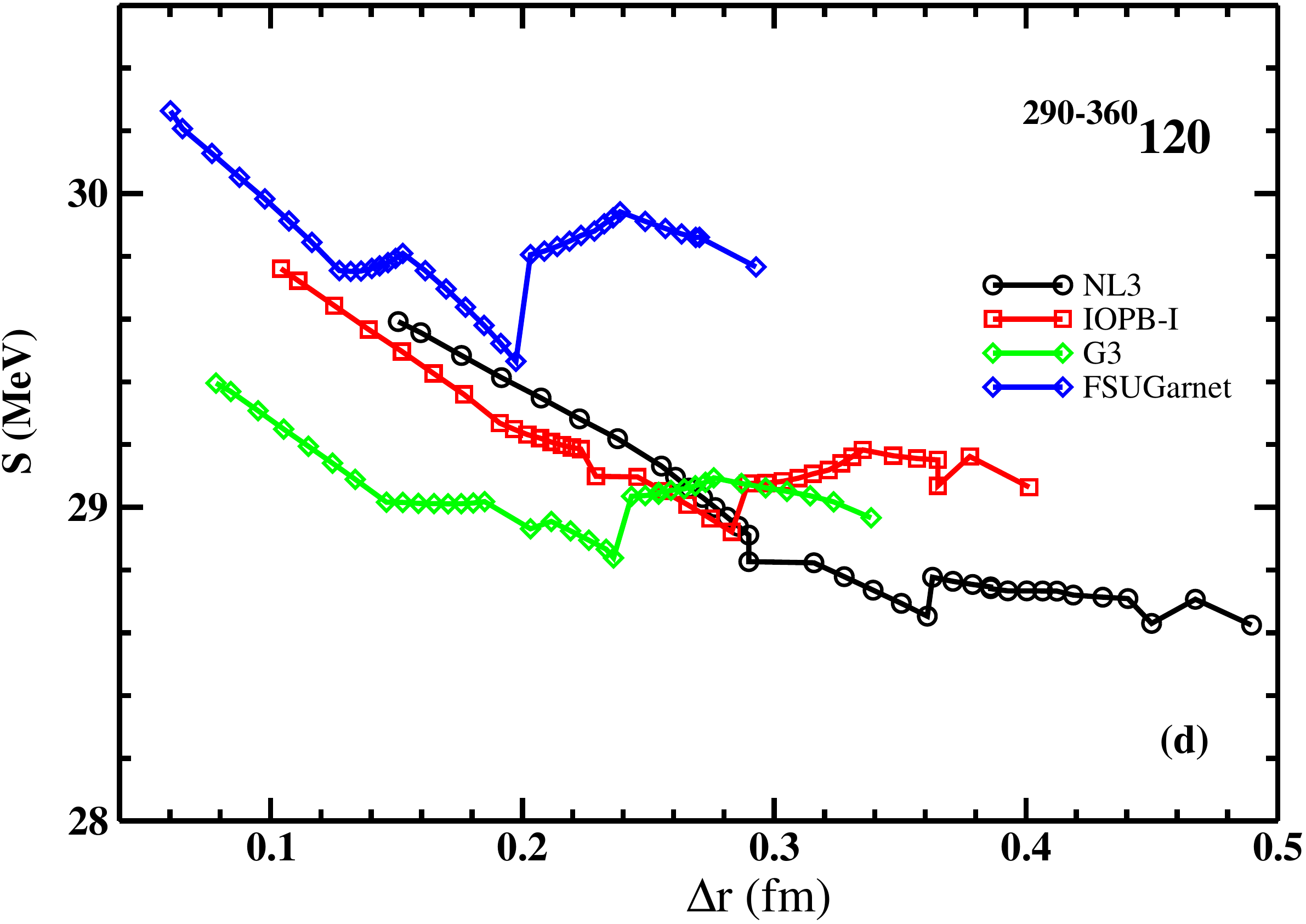}
        \caption{(color online) The correlation of the neutron skin-thickness with the symmetry energy is 
shown for the isotopes of Ca, Sn, Pb,  and Z = 120 nuclei as the representative cases corresponding to NL3, 
IOPB-I, G3, and FSUGarnet parameter sets. }
        \label{drS}
\end{figure*}

\subsection{The effective surface properties of the nuclei} 
\label{surface}
Before discussing the results of the symmetry energy and its derivatives, it is important to brief why we have 
considered spherical densities of the isotopes despite the fact that some of the isotopes, considered here, might be deformed. 
Generally, the symmetry energy coefficient of a finite nucleus is a bulk property, which mainly depends on the 
isospin asymmetry of the nucleus. Further, the symmetry energy coefficient of a nucleus can be written in terms 
of the volume and surface-symmetry energy coefficients (see Eq. 2 of Ref. \cite{bk12}). It is remarked in \cite{niko11} that 
the volume-symmetry energy is shape-independent, so the surface effects do not play their role in the volume-symmetry energy. 
On the other hand, the surface effects become negligible for heavy and superheavy nuclei since the surface symmetry 
energy coefficient is proportional to $A^{-1/3}$ , where $A$ is the mass number \cite{bk12,niko11}. It is clear from the previous 
sentence that surface effects are important for lighter nuclei and have negligible effects for heavier nuclei. 
The inclusion of the deformation may slightly improve the results but the effects of deformation are very small. 
In some of the recent works \cite{qiu15}, authors have shown the effect of deformation on the symmetry energy of finite nuclei 
using Thomas-Fermi approximation over Skyrme energy density functional. 
Following the work, one can find the relative change in the symmetry energy with very large deformation ($\beta_2 \sim 0.6$) 
is around 0.4 MeV. It is mentioned therein \cite{qiu15} that the effects of deformation decrease with respect to the mass number. 
So for the sake of computational ease, we have considered the spherical densities of the isotopes.

The effective surface properties for the isotopic series of O, Ca, Ni, Zr, Sn, and Pb nuclei are shown in 
Fig. \ref{spkn}. The first, second, and third row of each panel of the figure represent the symmetry energy
$S$, neutron pressure $P$, and symmetry energy curvature $\Delta K$, respectively. The value of the 
symmetry energy for finite nuclei lie in the range of 24-31 MeV. It is observed from the figure that the 
symmetry energy is larger for the FSUGarnet parameter set for all the cases except the isotopes of O and few isotopes of Ca. 
While G3 set predicts less symmetry energy in most of the cases in Fig. \ref{spkn}. The nature of 
parameter sets get reversed in the cases of neutron pressure $P$ and symmetry energy curvature $\Delta K$. 
For example, the G3 parameter set predicts the larger value of $P$ and $\Delta K$ for all the isotopic series except 
$N = 138$ isotope of Pb. 
Furthermore, we find several peaks at neutron numbers, which correspond to the magic numbers and/or shell
/sub-shell closures for each isotopic chain. These peaks in the symmetry energy curve imply that the 
stability of the nuclei at the magic neutron number is more as compared to the neighboring isotopes. 
The peaks in the symmetry energy curve imply that 
more energy would be required to convert one neutron to proton or vice verse. Apart from the peaks of the 
symmetry energy at the magic neutron number, a few small peaks are also evolved which may arise due to the 
shell structure on the density distribution of the nuclei. The present investigation predicts a few neutron 
magic numbers beyond the known magic numbers based on the well-known feature of symmetry energy over an 
isotopic chain, which will be studied systematically in the near future.  

The neutron pressure $P$ and symmetry energy curvature $\Delta K$ have the opposite nature to that of the 
symmetry energy with respect to the force parameter sets. It is meant by the opposite nature that higher 
the symmetry energy of nuclei corresponding to the particular interaction, lower the neutron pressure and 
symmetry energy curvature values are for the same parameter set and vice verse. Further, we found negative 
neutron pressure for the isotopes of Oxygen nuclei for all parameter sets. Also negative values of $P$ 
are obtained for the isotopic series of Ca and Ni corresponding to the FSUGarnet parameter set and a few of 
them corresponding to IOPB-I set. 
However, the NL3 set predicts negative values of $P$ for some of the isotopes of Ca and neutron-rich isotopes 
of Ni. It is to note that the negative value of $P$ arises due to the significant value of weight function 
(reflection of the behaviour of density distribution) in the 
range of local coordinate x (fm), where the pressure of nuclear matter is negative. For example, in Fig. 
\ref{nms} (a), the red arrow bar represents the range of x wherein the pressure is negative. In this range 
of x, the weight function has non-zero definite value (see Fig. \ref{weight} (a)), which is when multiplied 
by the pressure of nuclear matter (in Eq. \ref{p0}), results in the negative pressure of a nucleus. The non-zero 
definite values of the weight functions in the lower range of x are obtained for lighter nuclei due to their 
small size. On the other hand, the weight functions have negligible values for heavier nuclei in the range 
of x wherein the pressure of nuclear matter is negative for the corresponding nuclei. In general, the 
pressure and symmetry energy curvature values increase with neutron number for an isotopic series while the 
symmetry energy decreases with the increase of neutron number.

Observing the behavior of effective symmetry energy, neutron pressure, and 
symmetry energy curvature, we find a kink and/or a fall at the magic (or shell closures) neutron numbers. In the case 
of the isotopic chain of Pb nucleus, we did not get a kink in the $S$ curve at N=126. In spite of that, there is a 
fall of the $S$ curve at N=126, beyond which it is almost constant for the few isotopes. This 
signifies that N = 126 as a weak magic number.
Moreover at N=126, the sharp fall of the $P$ and $\Delta K$ curves can be noticed from the figure as they are related 
with the first and second-order derivatives of the symmetry energy. 
Even the sharp kink is not observed in the $S$ curve at N=126 for the isotopic series of Pb likewise at the 
other neutron magic numbers in different isotopic series, the sharp fall in the $P$ and $\Delta K$ curves support 
N=126 as a magic number. Furthermore, in Fig. 5 of Ref. \cite{gai11}, one can find that there are small kinks/deviation 
in the $S$ and $P$ curves at $^{208}$Pb while sharp kinks are observed at the other neutron magic numbers in 
the isotopic series of Ni (Fig. 1) and Sn (Fig. 6) \cite{gai11}. 

The effective surface properties for the isotopic series of O, Ca, Ni, Zr, Sn, Pb nuclei motivate us to 
pursue the said calculations for isotopes of experimentally unknown superheavy nuclei. Recently, the analysis 
of the superheavy element is a frontier topic in nuclear physics. The discovery of transuranic elements from 
Z = 93$-$118 with Oganesson ($_{118}Og$) is the heaviest element known so far that completes the $7p$ 
orbitals. Hence, the next element Z = 119 will occupy a new row in the Periodic Table. A large number of 
models predict different neutron and proton combinations for the next double close magic nuclei in the 
superheavy stability valley \cite{bhuyan12}. Among them, Z = 120 attracts much attention with neutron number 
N = 184 as the next double magic isotope, and near to be synthesized. Therefore, we have calculated the 
effective surface properties for the isotopic chain of Z = 120 nuclei, shown in Fig. \ref{120spkn}. 
We found almost a smooth fall in the symmetry energy up to N = 210, and further, a very miniature growth 
appears up to N= 240 following the previous trends. In other words, there is a moderate decrease in the 
symmetry energy over the isotopic chain of Z = 120 with some exception for the neutron number 
212 $\leq$ N $\leq$ 238 (see Fig. \ref{120spkn}). Here, 
we also got peaks in the neutron-rich side at N = 212 and 238 in the symmetry energy curve as similar to 
those in Fig. \ref{spkn}. These neutron numbers can be attributed to the magic neutron numbers. Here also, the symmetry 
energy predicted by the FSUGarnet parameter set is larger compared to the rest of the parameter sets. The 
neutron pressure and symmetry energy curvature are shown, respectively, in the second and third rows of Fig. \ref{120spkn}. 
Similar conclusions can be drawn for the $P$ and $\Delta K$ of Z = 120 isotopes, as for other 
isotopic chains. In general, we get a bit larger values of the effective surface properties for Z = 120 
isotopes as compared to the isotopes of the rest of the nuclei except a few isotopes of Sn and Pb. 
Here, we did not find any transparent signature of shell closures or magicity in the symmetry energy over the isotopic 
chain of Z =120. Following Ref. \cite{bhu12}, the ground state configuration of Z = 120 isotopes are super-deformed prolate and/or 
oblate shapes followed by a spherical intrinsic excited state. Here, our calculation limited to the spherical 
co-ordinate, which may cause for weaken the signature over the isotopic chain. Hence, a self-consistent 
microscopic calculation is required in the deformed basis.    

\subsection{Correlation of skin-thickness with the symmetry energy}
\label{correlation}
The skin-thickness is correlated with the surface properties  
\cite{gai11,bhu18,anto,gai12} for a different series of isotopic nuclei. It was found to be 
linearly correlated with the surface properties except for some kinks, which correspond to the magic/semi-magic 
nuclei of an isotopic chain \cite{gai11,bhu18,anto,gai12}. Here, we present the correlation between 
the symmetry energy and the neutron skin-thickness for the isotopic series of Ca, Sn, Pb, and Z = 120 nuclei 
for the NL3, IOPB-I, G3, and FSUGarnet parameter sets. It is remarked in Refs. \cite{aqjpg,iopb1} and shown 
in Table \ref{table3} that stiffer EoS of nuclear matter predicts the larger neutron skin-thickness of 
nuclei. Among the chosen parameter sets, NL3 is the stiffest which predicts larger skin-thickness, while 
FSUGarnet as being softer estimate smaller skin-thickness. On the other hand, it has been shown in Fig. 
\ref{spkn} that the symmetry energy is maximum at the neutron magic number of an isotope. 

Fig. \ref{drS} shows the correlation of the symmetry energy with the neutron skin-thickness of nuclei. The 
panels (a), (b), (c), and (d) of the figure represent the correlation for the isotopic series of Ca, Sn, Pb, 
and Z = 120, respectively. It is clear from the figure that the skin-thickness of the nuclei are larger 
corresponding to the NL3 parameter set and smaller for FSUGarnet set. It can be noticed from the figure that 
with some exceptions the symmetry energy 
predicted by FSUGarnet is higher compared to the rest of the parameter sets. The peaks in the symmetry energy 
curves (in Fig. \ref{drS}) correspond to the magic or semi-magic neutron numbers. The symmetry energy 
decreases with varying neutron numbers in either direction of the magic/semi-magic number. It implies that for  
exotic nuclei (nuclei lie at the drip-line) less amount of energy is required to convert one proton to 
neutron or vice-versa, depending on the neutron-proton asymmetry. The behavior of the symmetry energy with 
skin-thickness is undermined for a few cases. For the cases of Ca, Sn, and Pb, the symmetry energy curve is 
almost linear before and after the peaks. Further improvement in the results can be obtained by solving the 
field equations in an axially deformed basis. 

\section{Summary and Conclusions}
\label{summary}
In summary, we have studied the effective surface properties like the symmetry energy, neutron pressure, 
and the symmetry energy curvature for the isotopic series of O, Ca, Ni, Zr, Sn, Pb, and Z = 120 nuclei within 
the coherent density fluctuation model. We have used the spherically symmetric effective field theory 
motivated relativistic mean field model to study the ground state bulk properties of nuclei with the recent 
parameter sets like IOPB-I, FSUGarnet, and G3. The calculated results are compared with the predictions by 
the well known NL3 parameter set and found in good agreement. The densities of nuclei calculated within the E-RMF 
formalism are used as the inputs to the coherent density fluctuation model to obtain the weight functions 
for the isotopes. The symmetry energy, neutron pressure, and symmetry energy curvature of infinite nuclear 
matter are calculated within Brueckner energy density functional model which are further folded with the 
weight function to find the corresponding quantities of finite nuclei. The FSUGarnet parameter set predicts 
the large value of the symmetry energy while the smaller symmetry energy values are for G3 set 
with some exceptions. We found 
a larger value of the skin-thickness for the force parameter that corresponds to the stiffer EoS and vice-versa. 
We also found a few mass-dependence peaks in the symmetry energy curve corresponding to the neutron 
magic/semi-magic number. Observing the nature of the symmetry energy over the isotopic chain, we predict a 
few neutron magic numbers in the neutron-rich exotic nuclei including superheavy. The 
transparent signature of magicity is diluted for a few cases over the isotopic chain of Pb and Z = 120 
nuclei. Similar behavior is also observed for the neutron pressure and symmetry energy curvature for 
these isotopes. Concurrently, the present calculations tentatively reveal a way to calculate the effective 
surface properties of unknown drip-line nuclei including superheavy. More detail studies are disclosed by 
considering deformation into account. The calculated quantities are important for the structural properties of finite 
nuclei and may be useful for the synthesis of neutron-rich or superheavy nuclei. These effective surface properties 
can also be used to constrain an EoS of the nuclear matter and consequently nucleosynthesis processes.  

{\bf Acknowledgments:}
We are thankful to Shakeb Ahmad and Mitko Gaidarov for the fruitful discussions. AQ would like to 
acknowledge the Department of Science and Technology (DST), Gov. of India for providing financial support 
in the form of INSPIRE fellowship with No. DST/INSPIRE fellowship/2016/IF160131. 
The author (MB) has been supported by FAPESP Project Nos. 2014/26195-5 and 2017/05660-0, INCT-FNA Project Nos. 464898/2014-5 and 
by the CNPq -Brasil.

\end{document}